\documentclass[11pt,hyper,letterpaper]{JHEP3}
\usepackage{graphicx,amssymb,amsmath,amsfonts,epstopdf,bm}

\def\N{{\mathcal{N}}}

\def\<{\langle}
\def\th{\theta}
\def\>{\rangle}
\def\({\left (}
\def\){\right )}
\def\[{\left[}
\def\]{\right]}
\def\det{\rm{det}}
\def\beq{\begin{equation}}
\def\eeq{\end{equation}}

\def\jt{\langle J^t \rangle}
\def\jx{\langle J^x \rangle}
\def\jy{\langle J^y \rangle}
\def\jp{\langle J^+ \rangle}
\def\jm{\langle J^-\rangle}

\def\a{\alpha}
\def\s{\sigma}

\newcommand{\bea}{\begin{eqnarray}}
\newcommand{\eea}{\end{eqnarray}}

\def\ra{\rightarrow}
\def\lagr{{\cal L}}

\def\cp{\mathbb{CP}^2}
\def\nn{\nonumber}
\def\Eb{E_{\beta}}
\def\lag{\langle}
\def\rag{\rangle}
\def\Qp{\lag J^+ \rag}
\def\Qm{\lag J^- \rag}

\def\Qy{\lag J^y \rag}
\def\Qt{\lag J^t \rag}
\def\Qx{\lag J^x \rag}

\def\a{\alpha}
\def\b{\beta}

\def\g{\gamma}

\def\th{\theta}                   
\def\s{\sigma}                                   

\title{ \LARGE Holographic Flavor Transport in Schr\"{o}dinger Spacetime}
\author{Martin Ammon,$^1$\footnotemark[1]\, Carlos Hoyos,$^2$\footnotemark[2]\, Andy O'Bannon,$^1$\footnotemark[3]\, and Jackson M. S. Wu$^3$\footnotemark[4]
\\
$^1$Max-Planck-Institut f\"{u}r Physik (Werner-Heisenberg-Institut) \\ F\"{o}hringer Ring 6, 80805 M\"{u}nchen, Germany
\\
\\
$^2$Department of Physics, University of Washington \\ Seattle, WA 98195-1560, United States
\\
\\
$^3$Albert Einstein Center for Fundamental Physics \\
Institute for Theoretical Physics, University of Bern \\ Sidlerstrasse 5, 3012 Bern, Switzerland}

\footnotetext[1]{E-mail address: \email{ammon@mppmu.mpg.de}}
\footnotetext[2]{E-mail address: \email{choyos@phys.washington.edu}}
\footnotetext[3]{E-mail address: \email{ahob@mppmu.mpg.de}}
\footnotetext[4]{E-mail address: \email{jbnwu@itp.unibe.ch}}

\abstract{We use gauge-gravity duality to study the transport properties of a finite density of charge carriers in a strongly-coupled theory with non-relativistic symmetry. The field theory is $\N=4$ supersymmetric $SU(N_c)$ Yang-Mills theory in the limit of large $N_c$ and with large 't Hooft coupling, deformed by an irrelevant operator, coupled to a number $N_f$ of massive $\N=2$ supersymmetric hypermultiplets in the fundamental representation of the gauge group, \textit{i.e.} flavor fields. The irrelevant deformation breaks the relativistic conformal group down to the Schr\"odinger group, which has non-relativistic scale invariance with dynamical exponent $z=2$. Introducing a finite baryon number density of the flavor fields provides us with charge carriers. We compute the associated DC and AC conductivities using the dual gravitational description of probe D7-branes  in an asymptotically Schr\"odinger spacetime. We generically find that in the infrared the conductivity exhibits scaling with temperature or frequency that is relativistic, while in the ultraviolet the scalings appear to be non-relativistic with dynamical exponent $z=2$, as expected in the presence of the irrelevant deformation.}

\preprint{MPP-2010-37}
\begin{document}

\section{Introduction}

Gauge-gravity duality \cite{Maldacena:1997re,Gubser:1998bc,Witten:1998qj}, or more generally holography, provides a new tool for studying strongly-coupled systems at finite density, and in particular may provide novel insights into low-temperature systems controlled by quantum critical points. Quantum critical theories are invariant under scale transformations of the form
\beq
\label{eq:scaling}
t \rightarrow \lambda^z t, \qquad \vec{x} \rightarrow \lambda \, \vec{x},
\eeq
where $\lambda$ is some real, positive scaling parameter, $z$ is the ``dynamical exponent,'' and we have assumed spatial isotropy. If rotations, space translations, and time translations are also symmetries, then the theory is invariant under the so-called Lifshitz symmetry algebra. When $z=1$ the Lifshitz algebra may be enhanced to the relativistic conformal algebra.

The theory of fermions at unitarity, which can be realized experimentally using cold atoms, is invariant under the so-called Schr\"odinger symmetry~\cite{Mehen:1999nd,Nishida:2007pj}. The Schr\"odinger symmetry includes time translations, spatial translations, spatial rotations, and Galilean boosts, as well as scale transformations with $z=2$, a special conformal transformation, and a number symmetry with generator $N$, which is a central element of the Schr\"odinger algebra. The Schr\"odinger algebra is in fact easy to obtain from the relativistic conformal algebra in one higher spatial dimension. If we use the extra spatial dimension to form light-cone coordinates, $x^{\pm}$, and then retain only those generators that commute with the translation generator in the $x^-$ direction, $P_-$, the resulting algebra is precisely the Schr\"odinger algebra, if we make some identifications, including identifying the relativistic generator $P_+$ with the non-relativistic Hamiltonian (generator of time translations) and $P_-$ with the number operator $N$. Notice that if we want the spectrum of eigenvalues of $N$ to be discrete, which should be the case for a non-relativistic theory, then we must compactify $x^-$. In other words, if we begin with a relativistic conformal theory, break the symmetry group down to the subgroup that commutes with $P_-$ (via some deformation), and then perform a Discrete Light-Cone Quantization (DLCQ), then we obtain a non-relativistic theory with Schr\"odinger symmetry, in one lower spatial dimension.

Some gravitational duals for theories with Schr\"odinger symmetry were discovered in refs. \cite{Son:2008ye,Balasubramanian:2008dm} (see also refs. \cite{Duval:1990hj,Duval:2008jg}). Via the gauge-gravity dictionary, the Schr\"odinger symmetry group translates into the isometry group of the metric. We will thus call such spacetimes ``Schr\"odinger spacetimes.'' A direct method to obtain Schr\"odinger spacetimes, used in refs. \cite{Herzog:2008wg,Maldacena:2008wh,Adams:2008wt}, is to apply a solution-generating technique of type II supergravity, the Null Melvin Twist (NMT), to known solutions. We will review the NMT below. The basic example is to start with type IIB supergravity on $AdS_5 \times S^5$, where $AdS_5$ is (4+1)-dimensional anti-de Sitter space and $S^5$ is a five-sphere. The dual theory is $\N=4$ supersymmetric $SU(N_c)$ Yang-Mills (SYM) theory, in the limits of large $N_c$ and large 't Hooft coupling. Upon applying the NMT to this type IIB solution, we obtain $Sch_5 \times S^5$, where $Sch_5$ is (4+1)-dimensional Schr\"odinger spacetime. The solution also includes a non-trivial Neveu-Schwarz (NS) two-form, B. The dual theory is then $\N=4$ SYM deformed by a particular dimension-five operator that breaks the relativistic conformal group down to the Schr\"odinger group. We may then additionally compactify $x^-$, although doing so makes the supergravity approximation to string theory unreliable since $x^-$ is a null circle \cite{Maldacena:2008wh}. The generalization to thermal equilibrium states with temperature $T$ is straightforward \cite{Herzog:2008wg,Maldacena:2008wh,Adams:2008wt}.

As holographic models of fermions at unitarity, these systems have various advantages and disadvantages, some of which we review below. We will mention here two of the biggest disadvantages, however. First, the NMT does not produce a genuinely non-relativistic theory, but rather a deformation of a relativistic theory which then has Schr\"odinger symmetry. Second, as emphasized in ref. \cite{Adams:2009dm}, the $U(1)$ number symmetry generated by $N$ is not spontaneously broken in any of the known supergravity solutions, whereas real systems are typically superfluids.

Our goal is to study the transport properties of some ``charge carriers'' in such systems. For simplicity, we will work with the basic example above, type IIB supergravity in $Sch_5 \times S^5$. In the field theory, we will introduce a number $N_f$ of massive $\N=2$ supersymmetric hypermultiplets in the fundamental representation of the $SU(N_c)$ gauge group, \textit{i.e.} flavor fields. We will work in the probe limit, in which $N_f \ll N_c$, which amounts to ignoring quantum effects due to the flavors, such as the running of the coupling. The theory has a flavor symmetry analogous to the baryon number symmetry of Quantum Chromodynamics (QCD), with conserved current $J^{\mu}$. We will introduce a finite baryon number density, $\jt$, giving us our charge carriers. In condensed matter terms, we will study some dilute\footnote{Here dilute refers to energy density at finite temperature: the adjoint fields will have energy density of order $N_c^2$ while the flavor fields will have energy density of order $N_f N_c$.} massive charge carriers propagating through some quantum critical heat bath. We will then compute (holographically) both the DC and AC conductivities associated with baryon number transport. In the relativistic case these were computed in refs. \cite{Karch:2007pd,O'Bannon:2007in,Ammon:2009jt,Myers:2007we,Mateos:2007yp,Mas:2008jz}. The flavor fields appear in the supergravity description as a number $N_f$ of probe D7-branes in $Sch_5 \times S^5$. The mass and density are encoded in the D7-branes' worldvolume fields, as we will review.

Our study is complementary to that of ref.~\cite{Hartnoll:2009ns}, where the DC and AC conductivities of probe flavor were computed holographically using probe branes in Lifshitz spacetimes, that is, spacetimes whose isometry group is the Lifshitz group, with general $z$. Two of the main results of ref.~\cite{Hartnoll:2009ns} were that at temperatures low compared to the density and mass the DC conductivity $\sigma$ scales as $\jt T^{-2/z}$ (for all $z$) and the AC conductivity $\sigma(\omega)$ scales as
\beq
\label{eq:hpstaccondresults}
\sigma(\omega) \propto \left \{ \begin{array}{l l} \jt^{z/2} \, \omega^{-1} & \mbox{for $z<2$}, \\ \jt \left(\omega \log \omega \Lambda \right)^{-1} & \mbox{for $z=2$} \\ \jt \omega^{-2/z} & \mbox{for $z>2$}, \end{array} \right.
\eeq
(here $\Lambda$ is a dimensionful scale that renders the argument of the logarithm dimensionless). The authors of ref.~\cite{Hartnoll:2009ns} then suggested that, by introducing a scalar field, such as a dilaton, with nontrivial dependence on the holographic radial coordinate, the powers of $T$ and $\omega$ in the DC and AC conductivities, respectively, can be engineered to take essentially any value we like. In such a fashion we can produce holographic systems with scalings that match any number of real strongly-coupled electron systems. Moreover, with varying scalars we can even engineer flows in which the scalings change between the ultraviolet (UV) and infrared (IR). Examples of such flows, with exponent $z=2$ in the UV and $z=1$ in the IR, produced by relevant or marginally relevant deformations of Lifshitz spacetime, appear in refs. \cite{Kachru:2008yh,Cheng:2009df}.

Using Schr\"odinger instead of Lifshitz spacetime, we find that, in appropriate limits, for example low-temperature and large mass, the scalings with temperature or frequency in the IR are relativistic, meaning $z=1$, while in the UV the scalings are non-relativistic, meaning $z=2$. That is precisely what we expect in the dual (relativistic) field theory, in the presence of an \textit{irrelevant} operator that, roughly speaking, produces $z=2$ in the UV. Schr\"odinger spacetime is thus a good example of a flow from non-relativistic scaling in the UV to relativistic scaling in the IR.

Our results suggest that the NMT may be a useful tool for the kind of model-building proposed in ref.~\cite{Hartnoll:2009ns}. We can imagine starting with a \textit{relativistic} bulk system, introducing a scalar and engineering whatever exponents we like, and then performing a NMT. We will generically obtain a theory with Schr\"odinger symmetry, and exponents that flow from UV values, presumably with $z=2$, to the IR values we gave them in the original relativistic setting. Such an approach of engineering scaling exponents directly in a relativistic system may be technically easier than engineering them in a non-relativistic system.\footnote{AdS space is a solution of Einstein gravity plus a negative cosmological constant. Gravity alone cannot produce Lifshitz or Schr\"odinger spacetime, however. These require matter fields. In the Schr\"odinger case, the NMT takes AdS and generates the needed matter fields, in particular the NS two-form B. Introducing scalars in a theory of gravity alone and then doing the NMT may be easier than introducing scalars in a theory with gravity and other matter fields.}

The paper is organized as follows. In section \ref{sec:nmtandsch} we review how to add probe D7-branes to $AdS_5 \times S^5$, review the NMT and $Sch_5$ solution, and then discuss how the NMT affects the D7-branes' action, the Dirac-Born-Infeld (DBI) action. In section \ref{sec:dc} we compute the DC conductivity and in section \ref{sec:ac} we compute the AC conductivity. We conclude with some discussion and suggestions for future research in section \ref{sec:conclusion}.

\section{Adding Flavor to Schr\"odinger Spacetime}
\label{sec:nmtandsch}

In this section we will review how to obtain Schr\"odinger spacetime from a NMT of $AdS_5 \times S^5$, and review the field theory dual to supergravity on Schr\"odinger spacetime. Our new ingredient will be probe D7-branes. We will discuss in general terms what effect the NMT has on the D7-branes' worldvolume action.

\subsection{Review: D7-Branes in AdS}
\label{sec:review}

In type IIB supergravity, we begin with the solution describing the near-horizon geometry of non-extremal D3-branes, $AdS_5$-Schwarzschild times $S^5$. The metric is
\bea
\label{eq:d3metric}
ds^2 & = & g_{rr} \, dr^2 + g_{tt} \, dt^2 + g_{yy} \, dy^2 + g_{xx} \, d\vec{x}^2 + ds^2_{S^5} \\ & =& \frac{1}{r^2} \left( \frac{dr^2}{f(r)} -f(r) dt^2 + dy^2 + d\vec{x}^2 \right) + \left(d\chi + \mathcal{A} \right)^2 + ds^2_{\cp},
\eea
where $r$ is the radial coordinate, with the AdS boundary at $r=0$, and $(t,y,\vec{x})$ are field theory directions. We have singled out one field theory spatial direction, $y$, for use in the NMT below. Here $f(r) = 1 - r^4/r_H^4$, with $r_H$ the position of the black hole horizon. We are using units in which the radius of AdS is one, in which case the temperature of the black hole is $T=1/\left(\pi r_H \right)$. We have written the $S^5$ metric as a Hopf fibration over $\cp$, with $\chi$ the Hopf fiber direction. $\mathcal{A}$ gives the K\"ahler form $J$ of $\cp$ via $d\mathcal{A}=2J$. To write the metric of $\cp$ and $\mathcal{A}$ explicitly, we introduce $\cp$ coordinates $\a_1$, $\a_2$, $\a_3$, and $\theta$ and define the $SU(2)$ left-invariant forms
\bea
\label{eq:cp2coorddef}
\s_1 & = &  \frac{1}{2} \left( \cos \a_2 \, d\a_1 + \sin\a_1 \, \sin\a_2 \, d\a_3 \right), \nn \\ \s_2 & = & \frac{1}{2} \left( \sin \a_2 \, d\a_1 - \sin\a_1 \, \cos\a_2 \, d\a_3 \right), \nn \\ \s_3 & = & \frac{1}{2} \left( d\a_2 + \cos\a_1 \, d\a_3\right),
\eea
so that the metric of $\cp$ is
\beq
ds^2_{\cp} = d\theta^2 + \cos^2\theta \left( \s_1^2 + \s_2^2 + \sin^2\theta \, \s_3^2 \right),
\eeq
and $\mathcal{A} = \cos^2 \theta \, \s_3$. The full solution also includes a nontrivial five-form, but as shown in refs. \cite{Herzog:2008wg,Maldacena:2008wh,Adams:2008wt} that will be unaffected by the NMT, so we will ignore it.

When $T=0$, type IIB supergravity in the above spacetime is of course dual to $\N=4$ SYM theory in the limits of large $N_c$ and large 't Hooft coupling. The symmetries of the theory include the (3+1)-dimensional relativistic conformal group, dual to the isometry group of $AdS_5$, and an $SO(6)$ R-symmetry group, dual to the isometry group of the $S^5$. The $AdS_5$-Schwarzschild geometry is dual to the $\N=4$ SYM theory in a thermal equilibrium state with temperature $T$. In our units, we can convert from supergravity to SYM quantities using $\a'^{-2} = 4\pi g_s N_c = g_{YM}^2 N_c \equiv \lambda$, where $\a'$ is the string length squared, $g_s$ is the string coupling, $g_{YM}$ is the SYM theory coupling, and $\lambda$ is the 't Hooft coupling.

Following ref. \cite{Karch:2002sh}, we will introduce a number $N_f$ of probe D7-branes into the above geometry. The D7-branes will be extended along $AdS_5$-Schwarzschild times an $S^3 \subset S^5$. More specifically, the D7-branes will be extended along the three angular directions $\a_1$, $\a_2$ and $\a_3$. The two worldvolume scalars are then $\theta$ and $\chi$. As explained in ref.  \cite{Karch:2002sh}, the $N_f$ D7-branes are dual to a number $N_f$ of $\N=2$ supersymmetric hypermultiplets in the fundamental representation of the $SU(N_c)$ gauge group. We will work in the probe limit, which consists of keeping $N_f$ fixed as $N_c \rightarrow \infty$, so that $N_f \ll N_c$. On the gravity side, we may then ignore the back-reaction of the D7-branes on the supergravity fields. On the field theory side, we ignore quantum effects due to the flavor fields, such as the running of the coupling, as these are suppressed by powers of $N_f / N_c$.

Massless flavor fields break the $SO(6)$ R-symmetry to an $SO(4) \times U(1)$ symmetry, where the $U(1)$ and one $SU(2)$ subgroup of the $SO(4)$ form the R-symmetry of the $\N=2$ supersymmetric theory \cite{Kruczenski:2003be}. The $SO(4)$ symmetry is dual to the $SO(4)$ isometry of the $S^3$ that the D7-branes wrap. An $\N=2$ supersymmetric mass $m$ for the flavor fields preserves the $SO(4)$ but explicitly breaks the $U(1)$.

To study transport, we need to introduce a finite charge density, which we do as follows, following ref. \cite{Karch:2007pd}. We will introduce an $\N=2$ supersymmetry-preserving mass identical for all $N_f$ flavors. The field theory then has a global $U(N_f)$ flavor symmetry. In analogy with QCD, we will identify the overall diagonal $U(1)$ as baryon number (or more precisely, quark number). We will denote the associated conserved current as $J^{\mu}$, and study states with a finite density, that is, states with nonzero\footnote{We will always work in the canonical ensemble, with fixed $\jt$, rather than in the grand canonical ensemble, with fixed chemical potential.} $\jt$. To study transport we will introduce a constant external electric field that pushes on anything with $U(1)$ baryon number charge, pointing in the $x$ direction, \textit{i.e.} a constant $F_{tx}$. We will then compute (holographically) the resulting current $\jx$ and from that extract a conductivity.

Each field on the worldvolume of the D7-branes is dual to some gauge-invariant operator built from the flavor fields. The worldvolume field $\theta$ is dual to the mass operator  \cite{Karch:2002sh}, hence we will study embeddings with nonzero $\theta(r)$. The worldvolume scalar $\chi$ will be trivial, that is, the D7-branes will sit at a fixed value of $\chi$. Giving all the flavors the same mass means that in the bulk we introduce $N_f$ coincident D7-branes. The $U(N_f)$ gauge invariance of the D7-branes is dual to the $U(N_f)$ global symmetry of the field theory. The $U(1)$ worldvolume gauge field $A_{\mu}$ is dual to the $U(1)$ current $J^{\mu}$. To study states with nonzero $\jt$, $F_{tx}$, and $\jx$, we will introduce worldvolume gauge fields $A_t(r)$, $F_{tx}$, and $A_x(r)$.\footnote{Throughout the paper we use a gauge in which $A_r=0$.}

The D7-brane action describing the dynamics of the worldvolume fields is the DBI action plus Wess-Zumino (WZ) terms. For our ansatz with $\theta(r)$ and only the Abelian worldvolume gauge field, we will only need the Abelian D7-brane action. Additionally, for our ansatz the WZ terms vanish because the relevant form fields will not saturate the D7-branes' indices. In short, for our ansatz we only need the Abelian DBI action
\beq
\label{eq:generalDBIaction}
S_{D7} = - N_f T_{D7} \int d^8 \xi \, e^{-\Phi} \sqrt{-{\det} \left( P\left[g+B\right]_{ab} + \left( 2\pi\a'\right) F_{ab}\right)},
\eeq
where $T_{D7} = \frac{\a'^{-4} \, g_s^{-1}}{\left(2\pi\right)^7}$ is the D7-brane tension, the $\xi^a$ are the D7-branes' worldvolume coordinates (hence $a,b=0,1,\ldots7$), $\Phi$ is the dilaton, $P[g+B]_{ab}$ denotes the pull-back of the metric and NS two-form to the D7-branes' worldvolume, and $F_{ab}$ is the D7-branes' worldvolume $U(1)$ field strength.

Our ansatz for the gauge field involves only the $(r,t,x)$ directions, hence we may write $S_{D7}$ as a (3+1)-dimensional DBI action times some ``extra'' factors, with the (3+1)-dimensional part being the $(r,t,x,y)$ subspace:
\beq
\label{eq:4ddbi}
S_{D7} = - \N \int dr \cos^3\theta(r) \, g_{xx}^{1/2} \sqrt{-g - \frac{1}{2} g \tilde{F}^2 - \frac{1}{4} \left ( \tilde{F} \wedge \tilde{F} \right)^2 }.
\eeq
The square root factor is the characteristic form of a (3+1)-dimensional DBI action, in our case in the $(r,t,x,y)$ subspace, so that $g$ is the determinant of the induced metric in that subspace: $g = g_{rr}^{D7} g_{tt} g_{xx} g_{yy}$, where $g_{rr}^{D7}$ denotes a component of the induced metric on the D7-branes, so $g_{rr}^{D7} =  \frac{1}{r^2 f(r)} + \theta'(r)^2$. Starting now, primes will denote $\frac{\partial}{\partial r}$ and tildes will denote factors of $(2\pi\alpha')$, for example $\tilde{F}_{ab} = (2\pi\alpha') F_{ab}$. We have also performed the trivial integration over the $S^3$, producing a factor of the $S^3$ volume, $2\pi^2$, and defined
\beq
\label{eq:curlyndef}
\N \equiv N_f \, T_{D7} \, 2\pi^2 = \frac{1}{\left(2\pi\right)^4} \, \lambda \, N_f N_c,
\eeq
(not to be confused with the $\N$ of supersymmetry), where we have also written $\N$ in terms of SYM theory quantities. We also performed the trivial integration over the field theory directions and divided both sides by this (infinite) volume, so now $S_{D7}$ is actually an action density. We will use that convention in what follows. More explicitly, for our ansatz the DBI action is
\beq
\label{eq:adsd7action}
S_{D7} = - \N \int dr \cos^3\theta(r) \, g_{xx} \sqrt{|g_{tt}|g_{xx} g_{rr}^{D7} -\left(2\pi\alpha'\right)^2 \left( g_{xx} A_t'^2 + g_{rr}^{D7} \dot{A}_x^2 - |g_{tt}| A_x'^2 \right) },
\eeq
where dots denote $\frac{\partial}{\partial t}$. We define a Lagrangian $L$ via $S_{D7} = - \int dr L$.

The equations of motion for the gauge fields are trivial, since the action depends only on the derivatives $A_t'(r)$ and $A_x'(r)$. We thus obtain two constants of motion, which in terms of field theory quantities are simply $\jt$ and $\jx$. Explicitly, we have\footnote{For a rigorous derivation of $\langle J^{\mu} \rangle$, see ref. \cite{Karch:2007pd}.}
\beq
\label{eq:currentdefs}
\langle J^{\mu} \rangle = \frac{\delta L}{\delta A_{\mu}'}.
\eeq
These are two equations (for $\jt$ and $\jx$) for two unknowns ($A_t'(r)$ and $A_x'(r)$), hence we can solve just algebraically for $A_t'(r)$ and $A_x'(r)$. These solutions appear explicitly in ref. \cite{Karch:2007pd}.

We obtain $\theta(r)$'s equation of motion by varying the DBI action eq.~\eqref{eq:adsd7action} and then inserting the solutions for $A_t'(r)$ and $A_x'(r)$.  When the density is zero, $\jt=0$, but the temperature is finite, two topologically-distinct classes of embedding are possible \cite{Babington:2003vm,Kirsch:2004km}. In the first class, called ``Minkowski embeddings,'' at the boundary $r \rightarrow 0$ the D7-brane wraps the equatorial $S^3 \subset S^5$, but as the D7-brane extends into the bulk the $S^3$ shrinks and eventually ``slips off'' the $S^5$, collapsing to a point at some value of $r$, which we will call $r_{\Lambda}$, outside of the horizon (so $r_{\Lambda} < r_H$). From the AdS point of view, the D7-brane simply ends at $r_{\Lambda}$. In the other class of embeddings, called ``black hole embeddings,'' the $S^3$ shrinks but does not collapse, and the D7-brane intersects the horizon. As shown in ref.~\cite{Kobayashi:2006sb}, with a finite density $\jt$, such that $A_t(r)$ is non-trivial, only black hole embeddings are allowed.

We can extract the mass $m$ of the flavor fields from the coefficient of the leading term in $\theta(r)$'s asymptotic expansion, as explained in ref. \cite{Karch:2002sh}. Generically, solving the equation of motion for $\theta(r)$ and hence extracting the value of $m$, when the temperature and density are finite, requires numerics. Fortunately, we know the result for $\theta(r)$ in the two limits of zero and large mass. Zero mass corresponds to the trivial embedding, $\theta(r)=0$, in which case the D7-brane wraps the equatorial $S^3$ for all values of $r$. If we take the mass to be much larger than any other scale in the problem, then $\theta(r) \rightarrow \frac{\pi}{2}$. Here, if the density is zero, $\jt=0$, such that $A_t(r)$ is zero on the D7-brane worldvolume, then the D7-brane ends very close to the boundary, \textit{i.e.} $r_{\Lambda} \rightarrow 0$. At finite density, the D7-brane still has to reach the horizon. In that case the D7-brane forms a ``spike'' \cite{Kobayashi:2006sb}: the D7-brane \textit{almost} ends at some $r_{\Lambda}$, but then wraps a small $S^3$ of approximately constant volume all the way down to the horizon. In other words, $\theta(r)$ is nearly constant, almost but not quite equal to $\pi/2$, along the spike. In what follows we will either use the analytic solutions $\theta(r) = 0$ or $\theta(r) \rightarrow \pi/2$, or we will find approximate solutions for $\theta(r)$ in certain limits, as we do in section \ref{sec:ac}.

\subsection{The Null Melvin Twist}
\label{sec:nmt}

Following refs. \cite{Herzog:2008wg,Maldacena:2008wh,Adams:2008wt}, we now apply the NMT to the supergravity solution in eq.~(\ref{eq:d3metric}). The NMT is a species of TsT (T-duality, shift, T-duality) transformation that produces new supergravity solutions from old ones. The input is some solution with two commuting $U(1)$ isometries. The output is a new solution with different asymptotics. In our case, we will begin with $AdS_5$-Schwarzschild times $S^5$, using the Hopf fiber direction $\chi$ and the field theory spatial direction $y$ as isometry directions, and find a new solution for which the metric is asymptotically Schr\"odinger. The steps of the NMT are:
\begin{enumerate}
\item Boost by an amount $\g$ in $y$
\item T-dualize in $y$
\item Shift in the $\chi$ direction $d \chi \ra d \chi + \alpha dy$
\item T-dualize in $y$
\item Boost by $-\g$ in $y$
\item Take a limit: $\g \ra \infty$ and $\alpha \ra 0$ keeping $\b = \frac{1}{2} \alpha e^{\g}$ fixed
\end{enumerate}

Steps 2, 3, and 4, are the TsT part of the NMT. We will not write the explicit result of each step (for that, see ref. \cite{Adams:2008wt}), but we will make some generic comments about each step. The first step changes $g_{tt}$ and $g_{yy}$ and generates a $dy \, dt$ term in the metric. The second step produces a nontrivial NS B-field, with $dy \wedge dt$ component, and dilaton, and also changes $g_{tt}$ and $g_{yy}$. The third step changes $g_{yy}$ and also introduces a $d\chi + {\mathcal{A}}$ and $dy$ cross-term in metric, but leaves the B-field and dilaton unchanged. The fourth step changes $g_{tt}$, $g_{yy}$, $g_{ty}$, and the $\left(d\chi + \mathcal{A}\right)^2$ term in the metric, and generates a $\left(d\chi + {\mathcal{A}}\right)\wedge dy$ term in B. In the end, the metric that results from the NMT is asymptotically $Sch_5$. Explicitly, the final result for the metric is
\bea
\label{eq:schmetric}
ds^2 & = & \frac{1}{r^2} \left( \frac{dr^2}{f(r)} -f(r) \frac{1+ \b^2 r^{-2}}{K(r)} dt^2 + \frac{1-\b^2 r^{-2} f(r)}{K(r)} dy^2 - \frac{2 \b^2 r^{-2} f(r)}{K(r)} dt \, dy + d\vec{x}^2 \right) \nn \\ & & \qquad \qquad \qquad \qquad \qquad \qquad \qquad \qquad + \frac{1}{K(r)} \left(d\chi + \mathcal{A} \right)^2 + ds^2_{\cp}, \nn \\ & = & \frac{1}{r^2} \left( \frac{dr^2}{f(r)} - \frac{f(r)}{r^2 K(r)} dx^{+ 2} + \frac{2}{K(r)} dx^+ dx^- + \frac{1-f(r)}{2 K(r)} \left( \frac{dx^+}{\sqrt{2} \b} - \sqrt{2} \b dx^- \right)^2 + d\vec{x}^2 \right) \nn \\ & &  \qquad \qquad \qquad \qquad \qquad \qquad \qquad \qquad + \frac{1}{K(r)} \left(d\chi + \mathcal{A} \right)^2 + ds^2_{\cp},
\eea
where
\beq
f(r) = 1 - \frac{r^4}{r_H^4}, \qquad K(r) = 1 + \frac{\b^2 r^2}{r_H^4},
\eeq
and in the second equality of eq. (\ref{eq:schmetric}) we have introduced light-cone coordinates $X^{\pm}$,
\beq
\label{eq:bigxdef}
X^+ = t + y, \qquad X^- = \frac{1}{2}\left(- t + y \right),
\eeq
which we then rescaled by factors of $\beta$ to produce the light-cone coordinates $x^{\pm}$,
\beq
\label{eq:smallxdef}
x^+ = \b \left( t + y \right), \qquad x^- = \frac{1}{2\b} \left( -t + y \right).
\eeq
We discuss the utility of this rescaling at the end of this subsection. The solution also includes the NS two-form field
\bea
\label{eq:schbfield}
B & = & -\frac{\b}{r^2 K(r)} \left(d\chi + \mathcal{A} \right) \wedge \left( f(r) \, dt + dy \right), \nn \\ &= & - \frac{1}{2 r^2 K(r)} \left(d\chi + \mathcal{A} \right) \wedge \left( \left( 1 + f(r) \right) \, dx^+ + \left( 1 - f(r) \right) 2 \b^2 dx^- \right)
\eea
and a dilaton
\beq
\label{eq:schdilaton}
\Phi = - \frac{1}{2} \log K(r).
\eeq
Notice that if we take $\b \rightarrow 0$ then we recover the pre-NMT solution.

The metric in eq. (\ref{eq:schmetric}) has a horizon at $r=r_H$, with some associated Hawking temperature $T$. We obtain $T=0$ by sending $r_H \rightarrow \infty$, which sends $f(r) \rightarrow 1$ and $K(r) \rightarrow 1$. The resulting metric is then the metric of $Sch_5$ (not just asymptotically $Sch_5$), which is essentially the metric of $AdS_5 \times S^5$ with an extra term $-dx^{+2}/r^4$,
\bea
\label{eq:zeroTschmetric}
ds^2 & = & \frac{1}{r^2} \left( dr^2 - \frac{\beta^2}{r^2} dX^{+2} + 2 dX^+ dX^- + d\vec{x}^2 \right) + ds^2_{S^5} \nonumber \\ & = & \frac{1}{r^2} \left( dr^2 - \frac{1}{r^2} dx^{+2} + 2 dx^+ dx^- + d\vec{x}^2 \right) + ds^2_{S^5}
\eea
Notice that the $dx^{+2}$ term diverges faster as $r \rightarrow 0$ than the metric of $AdS_5$. When $T=0$ the geometry includes an $S^5$, however, the NS B-field breaks the $SO(6)$ isometry of the $S^5$ down to $SU(3) \times U(1)$, which is the isometry group of $\cp$. The $T=0$ solution also breaks all supersymmetry\footnote{Supersymmetric $Sch_5$ solutions do exist \cite{Bobev:2009mw,Donos:2009xc}, obtained by using different directions of the $S^5$ (besides the Hopf fiber) in the TsT transformation, which are very similar in form to the solution above. We will leave a thorough analysis of probe branes in those backgrounds for the future.} \cite{Maldacena:2008wh} and has a singularity at $r=\infty$ \cite{Adams:2008wt}. When $T$ is finite the geometry is only asymptotically $Sch_5$, the $S^5$ is deformed, the singularity is hidden behind a horizon, and the dilaton becomes non-trivial.

What is the field theory dual to type IIB supergravity on $Sch_5$? Equivalently, we can ask what field theory operation is dual to the NMT? Put briefly, the NMT is dual to adding an irrelevant operator to the $\N=4$ SYM theory Lagrangian. We can easily see this as follows. Given the solution above, if we perform a Kaluza-Klein reduction\footnote{For details of the reduction, which is in fact consistent, see ref. \cite{Maldacena:2008wh}.} on the $S^5$ the NS B-field gives rise to a massive vector in $Sch_5$ whose dual operator is a vector of dimension five, in the antisymmetric tensor representation of the $SU(4)$ R-symmetry. The dual operator, which we will denote $\mathcal{O}_{\mu}$, is a linear combination of operators of the form \cite{Herzog:2008wg}
\beq
\mathcal{O}_{\mu}^{IJ} = \textrm{Tr} \left( F_{\mu}^{~\nu} \Phi^{[ I} D_{\nu} \, \Phi^{J]} + \sum_K D_{\mu} \Phi^K \Phi^{[K} \Phi^I \Phi^{J]} \right) + \textrm{fermion terms},
\eeq
where $\Phi^I$ are the adjoint scalars of $\N=4$ SYM transforming in the \textbf{6} of $SU(4)$, $F_{\mu\nu}$ is the field strength, and $D_{\mu}$ is the covariant derivative. To be precise, recall that the \textbf{15} of $SU(4)$ decomposes into representations of $SU(3)$ as $\textbf{15} = \textbf{8} + \textbf{3} + \bar{\textbf{3}} + \textbf{1}$, so that we can write $\mathcal{O}_{\mu} = M_{IJ} \mathcal{O}^{IJ}_{\mu}$ where $M_{IJ}$ is an $SU(4)$ matrix that, after the decomposition, is in the \textbf{1} of $SU(3)$. In short, the NMT generates an NS B-field whose presence indicates an irrelevant deformation of $\N=4$ SYM: we have added $\mathcal{O}_+$ to the $\N=4$ SYM Lagrangian. Indeed, adding $\mathcal{O}_+$ breaks the relativistic conformal group down to the algebra of generators that commute with $P^+ \propto P_-$, producing the Schr\"odinger group, and breaks the $SU(4)$ R-symmetry down to $SU(3) \times U(1)$. As an irrelevant deformation, we also expect the geometry to be deformed near the boundary, which is indeed the case: we see explicitly in eq. \eqref{eq:zeroTschmetric} that the effect of the deformation (the $\beta^2$ term) grows near the boundary $r \rightarrow 0$.

The number generator $N$ of the Schr\"odinger algebra is dual to the isometry of the $x^-$ direction. If we want the eigenvalues of $N$ to be discrete, we must compactify $x^-$, that is, we must perform a DLCQ. When $T=0$ a DLCQ of the above geometry produces a null circle. Any closed strings that wrap the null circle will be massless, hence when $T=0$ and we compactify $x^-$ the supergravity approximation becomes unreliable \cite{Maldacena:2008wh}.  As emphasized in ref. \cite{Maldacena:2008wh}, if the spacetime has momentum in the $x^-$ direction then the $x^-$ circle is no longer null, and supergravity is reliable in most of the spacetime, although $x^-$ becomes null again near the boundary $r \rightarrow 0$.

The finite-$T$ solution in fact has $x^-$ momentum, since the NMT involves boosts, so with finite $T$ the $x^-$ direction is no longer null (as is obvious from eq. (\ref{eq:schmetric})). The dual field theory is in a state with a finite number density $N$, or equivalently a finite chemical potential \cite{Herzog:2008wg,Adams:2008wt}. As shown in refs. \cite{Herzog:2008wg,Adams:2008wt}, the field theory temperature $T$ and chemical potential\footnote{For us $\mu$ will always denote the chemical potential associated with the $U(1)$ along the compact $x^-$, not the chemical potential associated with the $U(1)$ baryon number of the flavor fields.} $\mu$ are
\beq
\label{eq:tandmudef}
T = \frac{1}{\pi r_H} \frac{1}{\beta}, \qquad \mu = - \frac{1}{2\b^2}.
\eeq
The dual field theory has the correct equation of state for a scale-invariant, non-relativistic theory with $z=2$ in two spatial dimensions (here we are performing a DLCQ), $\epsilon = P$, with $\epsilon$ the energy density and $P$ the pressure. The NMT does not change the area of horizons \cite{Gimon:2003xk}, so the metrics in eqs. (\ref{eq:d3metric}) and (\ref{eq:schmetric}) have the same horizon area, although the conversion to field theory quantities and interpretation differs in the two cases. Of central importance is the fact that the entropy, and other thermodynamic quantities, such as the free energy density, scale with \textit{negative} powers of $\mu/T$, and hence \textit{diverge} in the limit $\mu/T \rightarrow 0$. Such odd singular behavior appears to be a direct consequence of the DLCQ: exactly the same scalings occur in a gas of non-interacting, non-relativistic Kaluza-Klein particles \cite{Barbon:2009az}. Type IIB supergravity in Schr\"odinger spacetime is apparently not dual to a theory of fermions at unitarity.\footnote{Curiously, however, at $T=0$ and $\mu=0$, the three-point functions computed holographically from $Sch_5$ agree exactly, up to normalization, with the three-point functions of fermions at unitarity \cite{Fuertes:2009ex,Volovich:2009yh}. As in a relativistic conformal theory, the Schr\"odinger symmetry fixes the form of two-point functions but the three-point functions are only partially fixed, and so contain dynamical information. The fact that the $Sch_5$ result agrees with fermions at unitarity is thus a non-trivial statement about the dynamics of the dual theory.}

Notice that the bulk theory is relativistic, so that under a scale transformation the coordinates transform as $r \rightarrow \lambda r$, $t \rightarrow \lambda t$, $\vec{x} \rightarrow \lambda \vec{x}$ for some real positive number $\lambda$. The parameter $\beta$ has units of length and hence scales as $\beta \rightarrow \lambda \beta$. Rescaling the $X^{\pm}$ in eq. (\ref{eq:bigxdef}) by powers of $\beta$ produces the light-cone coordinates $x^{\pm}$ in eq. (\ref{eq:smallxdef}), such that under scalings $x^+ \rightarrow \lambda ^2 x^+$ while $x^-$ is invariant. Once we perform the DLCQ and interpret $x^+$ as the time coordinate, the resulting theory indeed exhibits the scaling of eq. (\ref{eq:scaling}) with $z=2$. The fact that $x^-$ is invariant indicates that the conjugate momentum $P_-$ is also invariant, which makes sense: after DLCQ we identify $P_-$ with the number operator $N$, which is a central element of the algebra, and in particular must commute with the dilation generator.

As reviewed for example in ref.~\cite{Hartnoll:2009ns}, for a theory with $d$ spatial dimensions, if we assign momentum to have scaling dimension one, then for a given value of $z$ we have the following scaling dimensions for a density $\jt$, current $\jx$, electric field $E$, magnetic field $B$, chemical potential $\mu$ and temperature $T$:
\beq
\label{eq:scalingdefs}
\left [ \jt \right ] = d, \quad \left [ \jx \right] = d + z - 1, \quad \left [ E \right ] = z+1, \quad \left [ B \right ] = 2, \quad \left [ \mu \right ] = \left [ T \right] = z.
\eeq
From Ohm's law, $\jx = \sigma E$, we find that the conductivity has dimension $\left [ \sigma \right ] = d-2$. Our system has $z=2$ and, after DLCQ, $d=2$. From eq. (\ref{eq:tandmudef}) we see that the factors of $\beta$ are essential to produce a $T$ and $\mu$ with scaling dimension two.

In the field theory we will also have background gauge fields, dual to the gauge fields on the D7-branes, such as the (relativistic) electric field $F_{tx}=\dot{A}_x$ in eq. (\ref{eq:adsd7action}). Here again, appropriate factors of $\beta$ will produce gauge fields with the correct scaling dimensions. From eq. (\ref{eq:tandmudef}) we have $\beta \propto \left(-\mu\right)^{-1/2}$, so we may interpret all rescalings by powers of $\beta$ as rescalings by appropriate powers of the chemical potential $\mu$. Recalling that the gauge field is a one-form, we have
\beq
\label{eq:rescaledgaugedef}
A_t dt + A_y dy = \frac{1}{2\beta}(A_t + A_y)dx^+ + \beta(-A_t + A_y)dx^- \equiv A_+ dx^+ + A_- dx^- \, ,
\eeq
so that $A_+$ has scaling dimension two while, after DLCQ, $A_-$ is a dimensionless scalar. An electric field $F_{+x} = \partial_+ A_x - \partial_x A_+$ will then indeed have scaling dimension $z+1=3$. Recalling the relativistic coupling $A_{\mu} J^{\mu}$, from the coupling $A_+ J^+$ we see that $J^+$ will have scaling dimension two, so that $A_+$ and $J^+$ have the correct scaling dimensions of a chemical potential and charge density, respectively, for $z=2$ and $d=2$. The coupling $A_- J^-$ indicates that after DLCQ $J^-$ will be a scalar with scaling dimension four.

\subsection{Twisting with Probe D7-branes}
\label{sec:nmtwithd7}

We now ask what happens to our probe flavor when we perform the NMT. The field theory side is easy, so we start there: we simply write the Lagrangian of $\N=4$ SYM theory coupled to massive $\N=2$ supersymmetric hypermultiplets in the fundamental representation of the gauge group, and then add the operator $\mathcal{O}_+$. The flavors break the $SU(3) \times U(1)$ symmetry to the same $SO(4) \times U(1)$ as in the relativistic case. In the probe limit, massless flavors will preserve the Schr\"odinger symmetry, while a finite mass will explicitly break scale invariance.

Now we ask what happens on the gravity side, that is, we ask what happens to probe D7-branes when we perform the NMT. The effect of the boost in steps 1 and 5 is straightforward. In the T-dualities of steps 2 and 4, the D7-branes are converted into D6-branes and then back to D7-branes. The component $A_y$ of the worldvolume gauge field is converted into a scalar, $\Phi_y$, which is then converted back into $A_y$. Crucially, however, the DBI action is consistent with T-duality \cite{Myers:1999ps}. That means that when we T-dualize, the metric, NS B-field, and dilaton may change, and we replace $A_y \rightarrow \Phi_y$, but the quantity
\beq
\lagr \equiv e^{-\Phi} \sqrt{{\det}\left(P\left[G+B\right]_{ab}+\left(2\pi\alpha'\right)F_{ab}\right)},
\eeq
evaluates to the same function of $r$, though now with $\Phi_y$ replacing $A_y$. If $A_y$ is non-trivial, so that after T-duality $\Phi_y$ is non-trivial, then the shift in step 3 may change the pullback of the metric to the D6-branes, and hence potentially change $\lagr$. If $A_y$ is trivial, however, then the TsT part of the NMT transformation leaves $\lagr$ unchanged.

The most general statement we can make is: if $\lagr$ is initially invariant under boosts in the $y$ direction, then the entire NMT has no effect on $\lagr$. In such cases the boosts in steps 1 and 5 and the TsT transformation each individually leave $\lagr$ unchanged. For boosts to be a symmetry requires $T=0$, and all worldvolume fields must be invariant under boosts in $y$.\footnote{In the Sakai-Sugimoto model \cite{Sakai:2004cn}, which is a system of intersecting D4-branes and D8-branes, the NMT seems to have no effect on the probe D8-branes' action, even in the black hole background and with nonzero $A_t(r)$ \cite{Pal:2008rf}.} For example, we may introduce the worldvolume scalar $\theta(r)$, as well as the gauge field $A_x(r)$, both of which are clearly invariant under boosts in $y$. In the field theory we will have flavor fields with a finite mass and some current in the $x$ direction, $\jx$.\footnote{\label{zeroTfootnote} As mentioned in ref. \cite{Karch:2007br} in the field theory at $T=0$ such a current will not dissipate. We may introduce it simply as an external parameter.} Using the asymptotically Schr\"odinger background and these worldvolume fields, we find that the action $S_{D7}$ is \textit{identical} to the asymptotically AdS case in eq. (\ref{eq:adsd7action}), with $A_t' = \dot{A}_x=0$. We may also add field strengths describing electric and magnetic fields pointing in the $y$ direction, such as $F_{ty}$, which are invariant under boosts\footnote{If we introduce an electric field $F_{ty}$ in the field theory, then we expect a resulting current in the $y$ direction, $\jy$, which breaks the boost invariance. A bulk solution with $F_{ty}$ and no $\jy$ would probably be pathological (exhibiting an instability of the kind that we will discuss in section \ref{sec:dcadsdlcq}, for example).}. The NMT leaves $\lagr$ invariant in all such cases.

A number of conclusions follow from the invariance of $S_{D7}$ for $y$-boost-invariant configurations. For example, suppose that, at $T=0$, we introduce only the worldvolume scalar $\theta(r)$. After the NMT we will find exactly the AdS result, eq. (\ref{eq:adsd7action}), with all gauge fields set to zero. The equation of motion for $\theta(r)$ is then identical to the AdS case, hence the solution is also identical: $\theta(r) = \arcsin(cr)$, where $c$ is a constant that determines the mass $m$ of the flavor fields via $m = c/(2\pi \alpha')$ \cite{Karch:2002sh}.\footnote{In AdS the $\theta(r) = \arcsin(cr)$ solution is supersymmetric, but here the background breaks all supersymmetry already, so we need not bother checking the supersymmetry of the D7-branes' embedding. A good question, though, is whether supersymmetric embeddings could be found for the supersymmetric Schr\"odinger solutions of refs. \cite{Bobev:2009mw,Donos:2009xc}.} The counterterms written in ref. \cite{Karch:2005ms}, needed to render the on-shell action finite, are then also identical to the relativistic case. Furthermore, in the $T=0$ AdS case, solutions with nonzero $\theta(r)$ and $A_x(r)$ were found in ref. \cite{Karch:2007br}. These solutions will also be identical for D7-branes in $Sch_5$.

Given that the embedding of the D7-branes, $\theta(r)$, is identical in the $T=0$ AdS and Schr\"odinger cases, a natural question is whether the spectra of linearized fluctuations of worldvolume fields are also the same. These spectra are dual to the spectra of mesons in the field theory. In general, the spectrum of mesons will not be the same. The simplest way to see that is to consider the fermionic mesons, dual to fermionic fluctuations of the D7-branes \cite{Kirsch:2006he}. The linearized equation of motion for these fermionic fluctuations is simply the Dirac equation. The Dirac operator is different in AdS and Schr\"odinger spacetimes \cite{Leigh:2009ck}. More generally, the differential operators appearing in the fluctuations' equations of motions, for example the scalar Laplacian, will differ from their AdS counterparts, so the spectrum will generically be different. Some subsector of the meson spectrum may be unchanged, for example the sector with zero momentum in $x^-$ and zero charge under the R-symmetry that corresponds to the Hopf fiber isometry. We leave a detailed investigation of the meson spectrum for the future.

When $T$ is nonzero the NMT changes $\lagr$, even when all the worldvolume fields are trivial. Introducing non-trivial worldvolume fields will then obviously not restore $\lagr$ to its AdS form. In what follows, we will be interested in finite $T$ solutions with worldvolume fields that are not invariant under boosts in $y$, such as electric fields $F_{tx}$, so we will not be able to exploit the $T=0$ invariance of $\lagr$.

In what follows we will study transport. We \textit{should}, however, first study thermodynamics, to determine the ground state of the system for all values of the parameters (mass, density, etc.). In the relativistic case, a variety of phase transitions do indeed occur as the parameters change (see refs. \cite{Babington:2003vm,Kobayashi:2006sb,Mateos:2006nu,Albash:2007bq,Erdmenger:2007bn,Erdmenger:2007ja,Erdmenger:2007cm,Erdmenger:2008yj} and references therein). We will leave a detailed analysis of the non-relativistic case for the future. In the following, in the field theory we will always assume that the D7-branes intersect the horizon, hence our results for the conductivity will only be valid when such D7-branes describe the ground state of the field theory.

\section{DC Conductivity of Probe Flavor}
\label{sec:dc}

In this section we will compute (holographically) a DC conductivity associated with transport of baryon number charge. We will use the method of ref. \cite{Karch:2007pd}, which captures effects beyond those of linear response. Our background spacetime will be asymptotically $Sch_5$ rather than asymptotically $AdS_5$. One of the major differences between these is that in $Sch_5$ we want to use light-cone coordinates $x^{\pm}$, compactify $x^-$, and in the dual non-relativistic theory interpret $x^+$ as the time coordinate. To understand how these operations affect the result for the conductivity, we first repeat the calculation of ref. \cite{Karch:2007pd} in the DLCQ of $AdS_5 \times S^5$ and then turn to the Schr\"odinger case.

\subsection{In the DLCQ of AdS}
\label{sec:dcadsdlcq}

Consider the trivial solution of type IIB supergravity: the metric is simply (9+1)-dimensional Minkowski space. Now introduce $N_f$ coincident probe D7-branes,\footnote{Strictly speaking, to avoid constraints on the number of D7-branes we should set the string coupling to be precisely zero. Many of the arguments that follow rely only on the form of the DBI action, rather than any properties unique to D7-branes, however.} and consider a solution in which the only nontrivial worldvolume field is a constant $U(1)$ electric field $F_{tx} = - E$. As in eq. (\ref{eq:4ddbi}), the DBI action assumes the characteristic (3+1)-dimensional form,
\bea
S_{D7} & = & - N_f T_{D7} \int d^8x \sqrt{1 - \frac{1}{2} \tilde{F}^2 - \frac{1}{4} \left( \tilde{F} \wedge \tilde{F} \right)^2} \nonumber \\ & = & - N_f T_{D7} \int d^8x \sqrt{1 - \left(2 \pi \alpha' \right) E^2}.
\eea
Clearly when the electric field is greater than the string tension, $E > 1/\left(2\pi\alpha'\right)$, the DBI action becomes imaginary. That signals the well-known tachyonic instability of open strings in an electric field \cite{Burgess:1986dw,Bachas:1992bh,Seiberg:2000zk}. The electric field pulls the endpoints of an open string in opposite directions. When the electric field is big enough to overcome the tension of the string, it rips the string apart. Another way to say the same thing is that the electric field reduces the effective tension of open strings. The instability appears when that effective tension becomes negative. Notice that if we additionally introduce a magnetic field $F_{xy}=B$ orthogonal to the electric field, then the DBI action becomes
\beq
S_{D7} = - N_f T_{D7} \int d^8x \sqrt{1 - \left(2 \pi \alpha' \right) \left( E^2 - B^2\right)}.
\eeq
If $E>B$ we can boost to a frame in which the magnetic field vanishes, and the arguments above still apply. If $E \leq B$ then the instability never appears. Indeed, in that case we can boost to a frame where the electric field is zero.

Now instead of flat space consider $AdS_5$-Schwarzschild times $S^5$, as in eq. (\ref{eq:d3metric}). Here the effective tension of strings already decreases as a function of $r$, going to zero at the horizon. Probe D7-branes with a constant worldvolume electric field will reduce the effective tension by the same amount at every value of $r$. We thus expect that for \textit{any} nonzero $E$ the effective tension will go to zero at some radial position $r_*$ outside the horizon, and to be negative between $r_*$ and the horizon. With an asymptotically AdS space, however, we have a dual field theory, so we can use our field theory intuition to guess the endpoint of the instability. The endpoint of a string looks like a quark. The electric field ripping a string apart should look like a Schwinger pair-production process. We should thus see a current. These arguments lead us to the ansatz of section \ref{sec:review}, for which the DBI action appears in eq. (\ref{eq:adsd7action}).

The observation of ref. \cite{Karch:2007pd} was that, with $\dot{A}_x = - E$, the action in eq. (\ref{eq:adsd7action}) depends only on $r$ derivatives of $A_t$ and $A_x$, hence the equations of motion for the gauge fields are trivial. We obtain two ``constants of motion,'' that is, $r$-independent quantities, the currents $\jt$ and $\jx$ in eq. (\ref{eq:currentdefs}). If we think of eq. (\ref{eq:currentdefs}) as two equations for two unknowns, $A_t'$ and $A_x'$, then we may solve algebraically for the unknowns. We thus obtain solutions for $A_t'$ and $A_x'$ in terms of $E$, $\jt$, $\jx$, metric components, and $\theta(r)$. These solutions appear explicitly in ref. \cite{Karch:2007pd}. Plugging these back into the action, we obtain
\beq
S_{D7} \propto - \int dr \cos^6\theta(r) g_{xx}^{5/2} |g_{tt}|^{1/2} g_{rr}^{1/2} \sqrt{\frac{|g_{tt}| g_{xx} - \left( 2\pi\alpha'\right)^2 E^2}{\N^2 \left(2\pi\alpha'\right)^2 |g_{tt}| g_{xx}^3 \cos^6\theta(r) + |g_{tt}| \jt^2 - g_{xx} \jx^2}},
\eeq
where we use the notation of section \ref{sec:review} and drop some constant prefactors. We are assuming that the D7-brane intersects the $AdS_5$-Schwarzschild horizon, as must be the case when $\jt$ is nonzero, hence the $r$ integration runs from the horizon to the boundary. We can see the instability as follows. Consider the fraction under the square root. At the horizon $|g_{tt}| =0$, so the numerator is negative. At the boundary $r\rightarrow \infty$, $|g_{tt}| g_{xx}$ diverges as $r^4$, hence for any finite $E$ the numerator becomes positive. The numerator must have a zero in between, which in fact occurs precisely at $r_*$,
\beq
\label{eq:rstardef}
\left [ |g_{tt}| g_{xx} - \left( 2 \pi \alpha' \right) E^2 \right]_{r_*}= 0.
\eeq
Notice that when $E=0$, the above equation implies $r_* = r_H$, and that as $E$ increases, $r_*$ moves closer to the boundary. In other words, with a larger $E$ we can probe UV physics ($r_*$ is close to the boundary) while with a small $E$ we probe IR physics ($r_*$ is close to the horizon).

We also included a current in our ansatz, which allows the denominator under the square root to have a similar sign change: at the horizon the denominator is negative, but at the boundary it is positive. We are thus able to avoid an imaginary DBI action (to avoid the instability) if we demand that the denominator and numerator change sign at the same place, $r_*$. We thus require
\beq
\label{eq:denomsignchange}
\left [ \N^2 \left(2\pi\alpha'\right)^2 |g_{tt}| g_{xx}^3 \cos^6\theta(r_*) + |g_{tt}| \jt^2 - g_{xx} \jx^2 \right]_{r*} = 0.
\eeq
The value of $E$ determines $r_*$ via eq. (\ref{eq:rstardef}), and eq. (\ref{eq:denomsignchange}) then determines the unique value of $\jx$ that prevents the instability.\footnote{From a field theory point of view, we choose the mass, temperature, charge density, and the electric field, and the dynamics of the theory then determines the system's response, \textit{i.e.} the resulting current.} Converting to field theory quantities, we find $\jx = \sigma E$, where
\beq
\label{eq:adssigma}
\sigma = \sqrt{\frac{N_f^2 N_c^2 T^2}{16\pi^2} \sqrt{e^2 + 1} \cos^6\theta(r_*) + \frac{d^2}{e^2 + 1}},
\eeq
where
\beq
e = \frac{E}{\frac{\pi}{2} \sqrt{\lambda} T^2}, \qquad d = \frac{\jt}{\frac{\pi}{2} \sqrt{\lambda} T^2}.
\eeq
To complete our solution, we must solve numerically for the final worldvolume field, $\theta(r)$. That has been done in refs. \cite{Mas:2008ks,Mas:2009wf}. When we compute the DC conductivity we will always use analytic solutions in the limits of zero and large mass explained at the end of section \ref{sec:review}.

The result in eq. (\ref{eq:adssigma}) consists of two terms adding in quadrature. We can easily identify the physical origin of each term, following refs. \cite{Karch:2007pd,O'Bannon:2007in,Ammon:2009jt,Karch:2008uy}. The second term, proportional to $d^2$, describes the contribution to the current from the charge carriers we introduced explicitly via a nonzero $\jt$. The first term, proportional to $\cos^6\theta(r_*)$, describes the contribution to the current from charge-neutral pairs. The microscopic process producing these pairs is not immediately clear. As in ref. \cite{Karch:2008uy}, however, we can take a limit with $\jt=0$, $T=0$, $m=0$ (which means $\cos \theta(r_*) = 1$), in which case we find a finite conductivity $\sigma \propto \sqrt{E}$. In that limit, the electric field is the only scale in the problem, hence the pair production must occur via a Schwinger process. When $T$ is finite, however, thermal pair production may also occur. The first term preumably knows about both kinds of pair production.

Notice also that the result for the conductivity depends explicitly on the electric field, and may also have implicit dependence through $\cos \theta(r_*)$. In the regime of linear response we expect, essentially by definition of \textit{linear} response, the current to be linear in $E$ and hence the conductivity to be independent of $E$. Here we are capturing effects beyond linear response. Ultimately we can do so because the DBI action sums all orders in $\left(2\pi\alpha'\right) F_{ab}$. Translating to field theory quantities, that means the result for $\sigma$ actually accounts for all orders in $E/\sqrt{\lambda}$.

Now suppose we want to perform a DLCQ both in the bulk and in the field theory. To do so, we first write the AdS part of the metric in eq. (\ref{eq:d3metric}) in the light-cone coordinates $X^{\pm}$ of eq. (\ref{eq:bigxdef}),
\bea
ds^2 & = & g_{rr} dr^2 + g_{++} dX^{+2} + g_{--} dX^{-2} + 2g_{+-} dX^+ dX^- + g_{xx} d\vec{x}^2 \nonumber \\ & = & \frac{1}{r^2} \left( \frac{dr^2}{f(r)} + \frac{1}{4} \left(1-f(r)\right) dX^{+2} + \left(1-f(r)\right) dX^{-2} + \left( 1+ f(r)\right) dX^+ dX^- + d\vec{x}^2 \right) \nonumber,
\eea
Notice that when $T=0$ and $f(r)=1$, the metric, and its inverse, in the light-cone directions is strictly off-diagonal, $g_{++} = g_{--} = 0$ and $g^{++} = g^{--} = 0$.

After the DLCQ, we interpret $X^+$ as the new time coordinate. The boundary value of the D7-brane worldvolume field $A_+$ acts as a source for the field theory operator $J^+$. In the DLCQ we assume all physical quantities are independent of $X^-$, that is, that $\partial_-$ acting on any quantity gives zero. After the DLCQ, the relativistic equation for conservation of the current, $\partial_{\mu} \langle J^{\mu} \rangle = 0$, reduces to $\partial_+ \jp + \partial_i \langle J^i \rangle = 0$, with $i$ the index for the spatial directions. We thus interpret $\jp$ as the charge density after DLCQ.

To study states in the field theory with finite $\langle J^+ \rangle$ our ansatz for the worldvolume fields will always include $A_+(r)$, or equivalently $F_{+r}(r) = - A_+'(r)$. Suppose for the moment we also introduce $A_-(r)$. The DBI action will then involve terms of the form
\beq
\frac{1}{2} g\tilde{F}^2 \supset g^{rr}_{D7} \, g^{++} A_+'(r)^2 + g^{rr}_{D7} \, g^{--} A_-'(r)^2 + 2 g^{rr}_{D7} \, g^{+-} A_+'(r) A_-'(r).
\eeq
Taking variational derivatives and using eq. \eqref{eq:currentdefs}, we see that a nontrivial $A_+'(r)$ not only produces a finite $\jp$ in the field theory but also a finite $\jm$. In other words, in the field theory, if we introduce $\jp$, we must introduce $\jm$. We will discuss the meaning of this, from the field theory point of view, shortly. Furthermore, when $T=0$ and $g^{++}=0$, a nontrivial $A_+'(r)$ produces \textit{only} a nonzero $\jm$. To obtain a nonzero $\jp$ at $T=0$, we will thus also introduce $A_-(r)$. Notice that if we return to the original coordinates, with just $A_t(r)$, then we obtain both $A_+(r)$ and $A_-(r)$.

We also want a constant electric field, which after DLCQ should be $F_{+x} = - E$. Introducing $F_{+x}$ alone doing does not produce an instability of the DBI action, however. When $T=0$, for example, the inverse metric component $g^{++}=0$, hence if we introduce only $F_{+x}$ then the DBI action does not depend on the electric field at all, since $g \tilde{F}^2 \propto g^{xx} g^{++} F_{+x}^2 = 0$. Converting back to the original coordinates reveals what is happening: $F_{+x}$ describes perpendicular electric and magnetic fields $F_{tx}$ and $F_{yx}$ of equal magnitude, such that $g \tilde{F}^2 \propto E^2 - B^2 = 0$. We will thus also introduce $F_{-x}$, in which case the DBI action depends on both $F_{+x}$ and $F_{-x}$, and exhibits the expected instability. Introducing both $F_{+x}$ and $F_{-x}$ is the same as introducing $F_{tx}$ and $F_{xy}$. Similarly to the story with $A_+(r)$ and $A_-(r)$, in what follows we will begin with $F_{tx}$ and then switch to light-cone coordinates.

In summary, our ansatz for the worldvolume gauge field will be identical to the relativistic case, with $F_{rt} = A_t'(r)$, $F_{rx} = A_x'(r)$ and constant $F_{tx}$, but converted to light-cone coordinates.

The bulk field $A_-$ is dual to the field theory operator $J^-$. Given that we will be working with a nontrivial $A_-(r)$ in the bulk and states with nonzero $\jm$ in the field theory, a natural question is, from the field theory point of view, what is $J^-$?

After the DLCQ, $A_-$ is a bulk scalar and $J^-$ is a scalar operator. To gain some intuition for the physical meaning of $A_-$ and $J_-$ after DLCQ, consider a complex scalar field in the relativistic theory that carries the $U(1)$ charge associated with the current $J^{\mu}$, which in our case means the scalars in the $\N=2$ hypermultiplet. Consider in particular the kinetic terms, written in light-cone coordinates and with a covariant derivative $D_{\mu} = \partial_{\mu} - i q A_{\mu}$ involving the background gauge field $A_{\mu}$, with $q$ the charge of the scalar under the $U(1)$. Explicitly, we have (here $\partial_{\pm}$ are derivatives with respect to $X^{\pm}$)
\bea
g^{\mu \nu} \left | D_{\mu}  \Phi \right |^{\dagger} D_{\nu} \Phi \nonumber & = & g^{+-}  \left | D_+  \Phi \right |^{\dagger} D_- \Phi + g^{-+}  \left | D_-  \Phi \right |^{\dagger} D_+ \Phi + \ldots \nonumber \\ & = & \partial_+ \Phi^{\dagger} \left( \partial_- - i q A_- \right) \Phi + \left( \partial_- + i q A_- \right) \Phi^{\dagger} \partial_+ \Phi + \ldots .\nonumber
\eea
If we work with fixed $X^-$ momentum $N$,
\beq
\Phi \left(X^-,X^+,\vec{x}\right) = e^{-i N X^-} \, \phi\left(X^+,\vec{x}\right),
\eeq
then we obtain
\beq
g^{\mu \nu} \left | D_{\mu}  \Phi \right |^{\dagger} D_{\nu} \Phi  = \left( N + q A_- \right) \, i \, \left [ \phi^{\dagger} \partial_+ \phi  - \left(\partial_+ \phi^{\dagger}\right) \phi\right] + \ldots.
\eeq
Recalling that after DLCQ we interpret $N$ as the ``particle number'' quantum number,\footnote{Here $N$ denotes the eigenvalue of the number operator, which we also called $N$ above.}, we see that a nonzero $A_-$ looks like a shift in the particle number $N$. Indeed, that will be true for \textit{any} $U(1)$: if we introduce a nonzero $A_-$, any fields charged under the $U(1)$ will appear to have a shifted $N$. That makes sense since, after DLCQ, a nonzero $A_-$ will produce a Wilson loop in the $x^-$ direction, effectively shifting the momentum $P_-$, and hence shifting $N$. Both $N$ and $q A_-$ couple to the operator representing $x^+$ momentum,
\beq
P_+ = i \left [ \phi^{\dagger} \partial_+ \phi  - \left(\partial_+ \phi^{\dagger}\right) \phi\right].
\eeq
As mentioned in the introduction (and ref.~\cite{Son:2008ye}), after the DLCQ, $P_+$ plays the role of the Hamiltonian in the Schr\"odinger algebra. We thus see that $A_-$ couples to the operator $J^- = q P_+$, or $q$ times the Hamiltonian. The statement above that a nonzero $\jp$ must be accompanied by a nonzero $\jm$ is thus easy to understand: from the perspective of the non-relativistic theory, a finite density of particles must be accompanied by some energy.

\subsection{In Schr\"odinger Spacetime}

Having explained the method for computing the DC conductivity, and some of the subtleties of working in light-cone coordinates, we proceed to the case where the background spacetime is the asymptotically $Sch_5$ metric of eq.~\eqref{eq:schmetric}.

First, we must be careful with factors of $\beta$ (see the end of section~\ref{sec:nmt}). As explained above, our ansatz is the same as in the relativistic case of section \ref{sec:review}, with $A_t(r)$, $F_{tx} = -E$, and $A_x(r)$, but converted to the (rescaled) light-cone coordinates $x^{\pm}$ of eq.~(\ref{eq:smallxdef}). We also rescale the gauge field components as in eq.~(\ref{eq:rescaledgaugedef}), so that our $A_{\pm}$ obey non-relativistic scaling. Explicitly, our ansatz for the worldvolume gauge field is
\beq
\label{eq:nonrelgaugefieldansatz}
A_+ (x,r) = \Eb \, x + h_+(r) \,, \quad A_- (x,r)= -2\beta^2\Eb \, x + h_-(r) \,, \quad A_x = A_x(r) \,,
\eeq
where we have redefined the electric field to be $\Eb = E/(2\beta)$ such that $\Eb$ scales non-relativistically, \textit{i.e.} $[\Eb]=3$, and $h_{\pm}(r)$ are functions for which we most solve. We also recall the other scaling dimensions of eq. (\ref{eq:scalingdefs}), with $d=z=2$,
\beq
[A_+] = 2, \quad [J^+] = 2, \quad [A_-]=0, \quad [J^-] = 4, \quad [A_x]=1, \quad [J^x] = 3,
\eeq
and the conductivity is dimensionless, $[\sigma]=d-2=0$.

We now want to insert our ansatz for the worldvolume fields into the DBI action, eq. \eqref{eq:generalDBIaction}, using the background metric, B-field, and dilaton of eqs.~\eqref{eq:schmetric}, \eqref{eq:schbfield}, and \eqref{eq:schdilaton}, respectively. To write the action succinctly, let us introduce some notation. We define
\beq
G_{i_1 \ldots i_n} \equiv \frac{1}{\sin^2\!\alpha_1} \, {\det}(P[g + B]_{ab}) \,, \quad a,b = i_1,\ldots,i_n \,,
\eeq
\textit{i.e.} $G_{i_1 \ldots i_n}$ is the determinant of the $n \times n$ submatrix of $P[g + B]_{ab}$ containing only rows and columns indexed by $i_1,\ldots,i_n$. All such sub-determinants are functions of $r$ times a factor of $\sin^2\a_1$, hence we divide by $\sin^2\a_1$ to make $G_{i_1 \ldots i_n}$ a function of $r$ only. Similarly we define the $3 \times 3$ submatrix determinant (divided by $\sin^2\a_1$):
\beq
G_B \equiv \frac{1}{\sin^2\!\alpha_1} \, {\det}
\begin{pmatrix}
 g_{+-}        & B_{+\alpha_2}        & B_{+\alpha_3} \\
-B_{-\alpha_2} & g_{\alpha_2\alpha_2} & g_{\alpha_3\alpha_2} \\
-B_{-\alpha_3} & g_{\alpha_2\alpha_3} & g_{\alpha_3\alpha_3}
\end{pmatrix} \,.
\eeq
Explicitly, the submatrix determinants we will need are
\begin{gather}
G_{\a_2\a_3} = \frac{\cos^2\!\theta(r) + K(r)\sin^2\!\theta(r)}{16K(r)}\cos^4\!\theta(r) \,, \notag \\
G_{+\a_2\a_3} = \frac{r^6 - 4r_H^4\beta^2 f(r)\sin^2\!\theta(r)}{64r^4 r_H^4\beta^2 K(r)}\cos^4\!\theta(r) \,, \qquad
G_{-\a_2\a_3} = \frac{K(r)-1}{16K(r)}\cos^4\!\theta(r) \,, \notag \\
G_B = \frac{1 + f(r)}{32r^2 K(r)}\cos^4\!\theta(r) \,, \qquad 
G_{+-\a_2\a_3} = -\frac{f(r)\cos^4\!\theta(r)}{16r^4 K(r)} \,.
\end{gather}
We note for later use that $G_{+-\a_2\a_3}$ vanishes at the horizon $r=r_h$ and otherwise is strictly negative, while $G_{\alpha_2\alpha_3}$ is strictly positive. We also define the shorthand notation
\begin{equation}
G_3 \equiv G_{-\alpha_2\alpha_3} + 4\beta^4 G_{+\alpha_2\alpha_3} + 4\beta^2 G_B 
= \frac{\beta^2[r^2 - \beta^2 f(r) \sin^2\!\theta(r)]}{4r^4 K(r)}\cos^4\!\theta(r) \,.
\end{equation}
Plugging our gauge field ansatz eq.~\eqref{eq:nonrelgaugefieldansatz} into the D7-brane action, we obtain
\begin{align}
\label{eq:schdccondaction}
S_{D7} &= -N_f T_{D7}\!\!\int\!dr\,d^3\a\,e^{-\Phi}\sqrt{-{\det}\left(P\left[g+B\right]_{ab}
+ \left(2\pi\a'\right) F_{ab}\right)} \notag \\
&= -\mathcal{N}\!\!\int\!\!dr \sqrt{-K(r)\,{\det}M_{ab}} \,,
\end{align}
where the matrix $M_{ab}$ has determinant
\begin{align}
{\det}M_{ab} \equiv g_{xx}\,g_{\a_1\a_1}\bigg\{
&g_{rr}\!\left[\Eb^2\,G_3 + g_{xx}G_{+-\alpha_2\alpha_3}\right] + G_{+-\alpha_2\alpha_3}(A_x')^2
+ \Eb^2\,G_{\alpha_2\alpha_3}(2\beta^2 A_+' + A_-')^2 \notag \\
+\,&g_{xx}\!\left[G_{-\alpha_2\alpha_3}(A_+')^2 + G_{+\alpha_2\alpha_3}(A_-')^2 - 2G_B A_+'A_-'\right]
\bigg\} \,.
\end{align}

Using eq. \eqref{eq:currentdefs} we obtain the current components
\begin{subequations}\label{eq:currents}
\begin{align}
\Qp &= \frac{\mathcal{N}K g_{xx}\,g_{\a_1\a_1}}{\sqrt{-K{\det}M_{ab}}}
\left[(2\beta^2\Eb^2\,G_{\alpha_2\alpha_3} - g_{xx}G_B)A_-' \,,
+ (4\beta^4\Eb^2\,G_{\a_2\a_3} + g_{xx}G_{-\a_2\a_3})A_+'\right] \, \\
\Qm &= \frac{\mathcal{N}K g_{xx}\,g_{\a_1\a_1}}{\sqrt{-K{\det}M_{ab}}}
\left[(2\beta^2\Eb^2\,G_{\a_2\a_3} - g_{xx}G_B)A_+'
+ (\Eb^2\,G_{\a_2 \a_3} + g_{xx}G_{+\a_2\a_3})A_-'\right] \,, \\
\Qx &= \frac{\mathcal{N}K g_{xx}\,g_{\a_1\a_1}}{\sqrt{-K{\det}M_{ab}}}\,G_{+-\a_2\a_3}A_x' \,.
\end{align}
\end{subequations}
Solving eqs.~\eqref{eq:currents} for $A_\pm'$ and $A_x'$ and plugging the solutions back into the action, we obtain the on-shell action
\beq\label{eq:SD7Sch}
S_{D7} = -\mathcal{N}^2\!\!\int\!\!dr K(r)g_{xx}g_{rr}^{1/2}g_{\a_1\a_1}
\sqrt{\frac{g_{xx}|G_{+-\a_2\a_3}| - \Eb^2\,G_3}{U(r) - V(r)}}
\eeq
where
\begin{subequations}
\begin{align}
U(r) &= \frac{\Qx^2}{G_{+-\a_2\a_3}} + \mathcal{N}^2 K(r)\,g_{xx}\,g_{\a_1\a_1} \,, \\
V(r) &= \frac{\Eb^2\,G_{\a_2\a_3}(\Qp - 2\beta^2\Qm)^2 
+ g_{xx}\left(G_{+\a_2\a_3}\Qp^2 + G_{-\a_2\a_3}\Qm^2 + 2G_B\Qp\Qm\right)} 
{g_{xx}G_{\a_2\a_3}(g_{xx}|G_{+-\a_2\a_3}| - \Eb^2\,G_3)} \,.
\end{align}
\end{subequations}

We now focus on the square root factor in the on-shell action in eq.~\eqref{eq:SD7Sch}, and demand that the action remain real for all $r$, as in the relativistic case of section~\ref{sec:dcadsdlcq}. First, notice that as a function of $r$, the factor in the numerator, $g_{xx}|G_{+-\a_2\a_3}| - \Eb^2\,G_3$, is negative at the horizon, $r = r_H$, and positive near the boundary $r \rightarrow 0$, and hence must have a zero at some $r = r_*$,
\beq
\label{eq:zeror}
\left[g_{xx}|G_{+-\a_2\a_3}| - \Eb^2\,G_3\right]_{r_*} = 0 \;\Longrightarrow\; 
\frac{1}{\Eb^2} = \frac{4r_*^2\beta^2}{f(r_*)}\left[r_*^2 - \beta^2 f(r_*)\sin^2 \theta(r_*)\right] \,.
\eeq
Similarly, the denominator $U(r) - V(r)$ is negative at the horizon and positive at the boundary, and so must have a zero also at some value of $r$. As in the relativistic $AdS$ case reviewed above, the two zeros must coincide to avoid an imaginary action, so we require $U(r_*) - V(r_*) = 0$. Now consider $V(r)$, which has a factor of $g_{xx}|G_{+-\a_2\a_3}| - \Eb^2\,G_3$ in its denominator. If the numerator of $V(r)$ is finite at $r_*$, then $V(r)$ will diverge at $r_*$. Notice, however, that $U(r)$ is not divergent at $r_*$. The only way to achieve $U(r_*) - V(r_*)= 0$ is thus to demand that the numerator of $V(r)$ vanish at $r_*$ (at least as quickly as $g_{xx}|G_{+-\a_2\a_3}| - \Eb^2\,G_3$). Setting the numerator of $V(r)$ to zero at $r_*$, we obtain (after some algebra)
\beq\label{eq:jmintermsofjp}
\Qm = -\frac{G_B + 2\beta^2 G_{+\a_2\a_3}}{2\beta^2 G_B + G_{-\a_2\a_3}}\bigg\vert_{r_*}\Qp  \,.
\eeq
Notice that eq.~\eqref{eq:jmintermsofjp} has no explicit dependence on the current $\jx$, and depends on the electric field only implicitly through $r_*$. In the absence of the electric field, eq.~\eqref{eq:jmintermsofjp} becomes $\Qm=-\Qp/(2\beta^2)$, which is independent of the temperature. Recalling the statements at the end of the last subsection, here we see that, indeed, we cannot introduce $\jp$ without also introducing $\jm$.

Now the condition $U(r_*) = V(r_*)$, combined with eq.~\eqref{eq:jmintermsofjp}, fixes the current to be
\begin{align}
\Qx^2 &= \Eb^2\left[\mathcal{N}^2 K g_{\a_1 \a_1}G_3
+ \frac{G_3^2\Qp^2}{g_{xx}^2(G_{-\a_2\a_3} + 2\beta^2 G_B)^2}\right]_{r_*} \\
&= \Eb^2\,\frac{f(r_*)}{64\,\Eb^2 \,r_*^6}
\left[\mathcal{N}^2 \cos^6 \theta(r_*) + \frac{16\Qp^2 r_*^2 f(r_*)}{\beta^4\Eb^2}\right] \,,
\end{align}
which gives the DC conductivity, $\sigma$, through Ohm's law $\Qx = \sigma\Eb$,
\beq
\sigma = \sqrt{\N^2 \frac{f(r_*)}{64\Eb^2r_*^6} \, \cos^6 \theta(r_*) +  \frac{ f(r_*)^2}{4 \beta^4 r_*^4 \Eb^4} \Qp^2}.
\eeq
Note that $\sigma$ is dimensionless, as it should be. Notice that once we fix $T$, $E_{\beta}$ and $\Qp$, reality of the action determines $\Qm$ and $\Qx$.

As in the relativistic case, the result for $\s$ consists of two terms adding in quadrature. Once again, the second term, proportional to $\Qp^2$, describes the contribution to the current from the charge carriers we introduced explicitly via the net density $\Qp$. The first term, again proportional to $\cos^6\theta(r_*)$, appears to describe the contribution from charge-neutral pairs. We suspect that, as in the relativistic case, these pairs come from Schwinger and/or thermal pair production. From the field theory point of view, such pair production at first seems counter-intuitive, since in a non-relativistic theory, the number of particles should not change. Recall, however, that we are actually studying a relativistic theory which we deform in two ways, first by introducing an irrelevant operator and then by performing a DLCQ. From that perspective, nothing is wrong with pair production. Notice also that what plays the role of the ``number of particles'' is the eigenvalue of $N \sim P_-$, the momentum in the $x^-$ direction, which is indeed fixed.

We now take two different limits to explore the scaling of the conductivity. The two limits depend on the relative strengths of $\Eb$ and $T$. In the limit of a very weak electric field, $\Eb^2\beta^2 r_H^4 \ll 1$, or equivalently $\Eb \ll \beta T^2$, we have from eq.~\eqref{eq:zeror} $r_* \ra r_H$, and the conductivity takes the form
\beq
\sigma \approx \sqrt{\frac{\pi^2\mathcal{N}^2 \cos^2\theta(r_*) }{16}T^2\beta^4 + \frac{4\Qp^2}{\pi^4\beta^4 T^4}}
= \sqrt{\frac{\pi^2\mathcal{N}^2 \cos^2\theta(r_*)}{64}\frac{T^2}{\mu^2}
+ \frac{16}{\pi^4}\frac{\Qp^2}{T^2}\frac{\mu^2}{T^2}} \,,
\eeq
where in the second equality we have replaced $\beta$ with $\mu$ using eq. \eqref{eq:tandmudef}. At low temperatures or large masses, \textit{i.e.} $T/\mu \rightarrow 0$ or $\cos \theta(r_*) \rightarrow 0$,  the pair-production term will be suppressed, in which case the conductivity approaches
\beq
\label{eq:largeT}
\sigma\approx\frac{4}{\pi^2}\frac{\Qp}{T}\frac{\left(-\mu\right)}{T} \,.
\eeq
In a scale-invariant theory with dynamical exponent $z$, the conductivity should behave as $\sigma \sim \Qp T^{-2/z}$. Here we see that, if we fix the chemical potential $\mu$, then we obtain the dynamical exponent of a relativistic theory, $z=1$. On the other hand, if we hold fixed the ratio $\mu/T$ as we vary the temperature, then we find non-relativistic scaling, with $z=2$.

The opposite limit is strong electric field, $\Eb^2\beta^2 r_H^4 \gg 1$, or equivalently $\Eb \gg \beta T^2$. If the mass is small, so that $\theta(r_*) \approx 0$ and hence $\sin \theta(r_*) \approx 0$, then eq.~\eqref{eq:zeror} implies that $r_* \sim (2\Eb\beta)^{-1/2}$, and we find
\beq
\label{eq:smallT}
\sigma\approx\sqrt{\frac{\mathcal{N}^2 \cos^6\theta(r_*)}{8}\Eb\beta^3 + \frac{\Qp^2}{\Eb^2\beta^2}}= \sqrt{\frac{\mathcal{N}^2 \cos^6\theta(r_*)}{8^{3/2}}\frac{\Eb}{\left(-\mu\right)^{3/2}} + 2\frac{\Qp^2\left(-\mu\right)}{\Eb^2}} \,.
\eeq

If we set the mass and density to zero, so $\cos \theta(r_*) = 1$ and $\Qp = 0$, then we obtain a finite conductivity,
\beq
\label{eq:dcsigmazeroT}
\sigma \approx  \frac{\mathcal{N}}{8^{3/4}}\sqrt\frac{\Eb}{\left(-\mu\right)^{3/2}},
\eeq
We indeed find a nonzero current, which must come from Schwinger pair production. The bulk mechanism is exactly the same as in the relativistic case: the worldvolume electric field is ripping strings apart. If we fix the value of $\mu$, then $\s \propto \sqrt{\Eb}$, which is the same scaling with the electric field as in the $T=0$ relativistic case. If we fix the ratio $\Eb/\left(-\mu\right)^{3/2}$ as we vary $\Eb$, however, then the conductivity is a constant, which we expect for a (2+1)-dimensional theory with non-relativistic scale invariance, if the only scale is the electric field.

If the electric field is small, then the second term under the square root in eq. \eqref{eq:smallT} dominates. We then find 
\beq
\label{eq:smallE}
\sigma \approx \frac{\Qp \left(-2 \mu\right)^{1/2}}{\Eb}\,,
\eeq
Switching back to relativistic coordinates,
\beq
\label{eq:qp}
\Qp = \beta (\Qt+\Qy) = \frac{1}{ \left(-2 \mu\right)^{1/2}} (\Qt+\Qy)\,,
\eeq
we find that, for a fixed $\mu$, the equation for the current becomes
\beq
\label{eq:currentLargeE}
\Qx = \Qt+\Qy\,,
\eeq
which is similar to the AdS result at zero temperature (see Appendix A of ref.~\cite{Karch:2008uy}). The physics here is simply that at zero temperature the charge carriers are accelarated to the speed of light, so the system is not really stationary. If we fix the ratio $\Eb/\left(-\mu\right)^{3/2}$, so that we can use $\left(-\mu\right)^{1/2} \sim E_{\beta}^{1/3}$, then
\beq
\label{eq:dcsigmazeroT2}
\sigma \propto \frac{\Qp}{\Eb^{2/3}}\,,
\eeq 
which is the appropriate dependence on the electric field for a two-dimensional non-relativistic conformal theory with a finite density. 

To summarize: of the two scales $E_{\beta}$ and $T$, we can take one to be large relative to the other. If we hold the larger scale fixed relative to the scale set by $\mu$, then we obtain non-relativistic scaling. If on the other hand we hold $\mu$ fixed and vary the larger scale, we obtain relativistic scaling.

We can understand our results in terms of the geometry as follows. Consider for example the case in which we can neglect the electric field, so the temperature is the larger scale. The conductivity is then evaluated at $r_*\simeq r_H$. Fixing $\mu/T$ is the same as fixing $r_H/\beta$. We are thus probing the geometry on the scale of the deformation $\beta$, so intuitively we expect to obtain non-relativistic behavior. Indeed, in this case the conductivity in \eqref{eq:largeT} behaves in a non-relativistic way. If instead we fix $\mu$ as we take $T/\mu \rightarrow 0$ (see the text above eq. \eqref{eq:largeT}), then the ratio $r_H/\beta\to \infty$ and the horizon enters the region where the geometry is similar to AdS-Schwarzschild. In that case \eqref{eq:largeT} indeed exhibits relativistic scaling.

\section{AC Conductivity of Probe Flavor}
\label{sec:ac}

We now proceed to compute the frequency-dependent conductivity in the linear response approximation. In the field theory we consider thermal equilibrium states with temperature $T$ and massive flavor fields with a finite density $\Qp$ and finite $\Qm$, but now, unlike the last section, no constant electric field.

In the holographic dual the system at equilibrium is described by probe D7-branes in the asymptoticaly Schr\"odinger black hole geometry of eq. \eqref{eq:schmetric}, with nontrivial worldvolume fields $A_+(r)$ and $A_-(r)$. We can obtain the solution for these in exactly the same way as the last section: each has an associated constant of motion, $\Qp$ and $\Qm$ from eq. \eqref{eq:currentdefs}, so we obtain equations similar to those in eqs. \eqref{eq:currents}, which we then algebraically invert to find $A_+(r)$ and $A_-(r)$ in terms of $\Qp$ and $\Qm$. We will not present these solutions explicitly, but we will record that, in the absence of a constant worldvolume electric field, we have from the last section that $r_*=r_H$ and $\Qm = -\Qp/(2\beta^2)$ (see eqs. \eqref{eq:zeror} and \eqref{eq:jmintermsofjp}). Notice also that, as always with a nontrivial $A_+(r)$ (or $A_t(r)$), the D7-branes must extend all the way to the black hole horizon \cite{Kobayashi:2006sb}. We will also consider a nontrivial embedding $\theta(r)$, whose form we will discuss in detail below.

To obtain the conductivity in the regime of linear response, we consider a small, frequency-dependent perturbation of the worldvolume electric field about the background solution (which has $A_+(r)$, $A_-(r)$, and $\theta(r)$),
\beq
\label{eq:axflucuation}
A_x(x^+,x^-,r) ={\rm Re}\left[\;e^{-i\omega (x^++2\beta^2 x^-)} a_x(r,\omega)\right]\,.
\eeq
For simplicity we will work with zero spatial momentum. Here we note that, in fact, removing all $x^-$ dependence does not qualitatively change our results for the behavior of the conductivity with frequency.

To quadratic order, the Lagrangian density for the perturbation is
\beq
\label{eq:actflucs}
\lagr_2 =\alpha_{++} f_{+x}^2+\alpha_{--}f_{- x}^2+2\alpha_{+-} f_{+x} f_{- x}-\alpha_{rr} f_{xr}^2,
\eeq
where $f_{\alpha\beta}=\partial_\alpha A_\beta-\partial_\beta A_\alpha$ is the field strength associated with the fluctuation in eq. \eqref{eq:axflucuation}. The equation of motion for the fluctuation is then
\beq
\label{eq:eomflucs}
a_x''+ \frac{\alpha_{rr}'}{\alpha_{rr}} \, a_x' + \omega^2\frac{\alpha_{++} + 4\beta^4\alpha_{--} + 4\beta^2\alpha_{+-}}{\alpha_{rr}} \, a_x = 0 \,.
\eeq
Here the $\a$ are $r$-dependent coefficients that depend on the background solutions for $A_+(r)$ and $A_-(r)$ about which we are perturbing. To write these succinctly, let us introduce some notation. We define
\beq
k_s(r) \equiv 1 + \frac{\sin^2 \theta(r) \, \beta^2 r^2}{r_H^4} \,, \qquad \rho(r) \equiv 1 + r^2 f(r){\th'}^2  \,,
\eeq
\beq
Q^2 \equiv \frac{64\Qp^2}{\beta^2} \,, \qquad \gamma^2(r) \equiv \cos^6\theta(r) k_s(r) + Q^2 r^4 (r^2+\beta^2 \sin^2\theta(r)) \,.
\eeq
The coefficients in the quadratic Lagrangian density are then
\begin{align}
\alpha_{rr}(r) & = \frac{\N}{16r\sqrt{k_s(r)\rho(r)}}f(r)\gamma(r) \,, \notag \\
\alpha_{++}(r) & = \frac{\N}{16r_H^4 f(r)\gamma(r)}\beta^2 r^3
\left(Q^2 r^2 r_H^4 + \cos^6\theta(r)\right)\sqrt{k_s(r)\rho(r)} \,, \notag \\
\alpha_{--}(r) & = \frac{\N}{64\beta^2 r^3 f(r)\gamma(r)}
\left[Q^2 r^8 + \cos^6\theta(r)\left(\frac{r^6}{r_H^4} - 4\sin^2\theta(r)\beta^2 f(r)\right)\right]\sqrt{k_s(r)\rho(r)} \,, \notag \\
\alpha_{+-}(r) & = \frac{\N}{32r f(r)\gamma(r)}
\left[Q^2 r^6 - \cos^6\theta(r)\left(\frac{r^4}{r_H^4} - 2\right)\right]\sqrt{k_s(r)\rho(r)} \,.
\end{align}

We follow the now-standard procedure to compute transport coefficients, in the regime of linear response, holographically (for a review see refs. \cite{Hartnoll:2009sz,Herzog:2009xv,McGreevy:2009xe}). We must first solve the linearized equation of motion for the fluctuation $a_x(r,\omega)$ with the boundary condition that near the horizon the solution has the form of a traveling wave moving into the black hole, \textit{i.e.} an in-going wave. We then insert that solution into the action, which then acts as a generating functional for field theory correlators. Taking two functional derivatives of the on-shell action gives us the retarded Green's function. We then extract the conductivity from the retarded Green's function via a Kubo formula. Ultimately, we find
\beq
\label{cond}
\sigma(\omega)\propto \lim_{r\to 0}\left[\frac{\alpha_{rr}a_x'(r,\omega)}{\omega a_x(r,\omega)}\right] \,,
\eeq
where, because we are primarily interested in $\sigma(\omega)$'s scaling behavior with $\omega$, we omit the overall prefactor.

Crucially, notice that the result will depend on the background solution $\theta(r)$ describing the embedding of the D7-brane. In general, with finite temperature $T$ and density $\Qp$, we can only solve for $\theta(r)$ numerically.  We leave a complete numerical solution for future work. Here we will focus on regimes of physical interest, applying some approximations to obtain analytic results for $\sigma(\omega)$'s scaling with $\omega$.

We are interested in a regime dominated by the physics of a zero-temperature ``critical point.'' Remember that the Schr\"odinger geometry interpolates between a UV critical point with $z=2$ and an IR critical point with $z=1$. The scale that separates the two regimes is the chemical potential ($\beta$ in the metric). In order to eliminate thermal effects, we will work in a limit where the temperature is much smaller than any other scale, which in particular means $\omega \gg T$. We then expect that we can explore both regions by taking  $\mu \gg \omega$ for the IR regime and $\mu \ll \omega$ for the UV.

In the probe approximation the critical behavior is not spoiled by a nonzero mass or charge density for the flavors. The sector described by the probe D7-branes is sensitive to these quantities, however, so we expect deviations from scale invariance whenever $\omega \sim \jp^{z/d}$ or $\omega\sim m^2$. In order to avoid such deviations we will only explore frequencies below these scales, which means $\jp^{z/d}\gg \mu$ and $m^2 \gg \mu$.

Instead of computing the exact current-current correlator in the holographic description, we will use the radial coordinate as an approximation to the frequency scale, following ref. \cite{Hartnoll:2009ns}. We will call $r_0$ the reference scale around which we will give an estimate of the conductivity. More precisely, we will define a ``local conductivity'' $\sigma(\omega,r_0)$ to be the quantity in brackets in eq.~\eqref{cond}, evaluated at $r=r_0$,
\beq
\label{eq:localcond}
\sigma(\omega,r_0) \equiv \left[\frac{\alpha_{rr}a_x'(r,\omega)}{\omega a_x(r,\omega)}\right]_{r_0}.
\eeq
If we think about the charge carriers as strings attached to the brane, the length of the string from $r_0$ to the horizon\footnote{We are using black hole embeddings only, in which case the endpoint of a string at $r_0$ would be free to move along the D7-brane and into the horizon. Our statements about the physical meaning of $r_0$ are meant only to provide intuition.} (times the string tension) is, roughly speaking, the energy of the charge carriers we are exciting when we apply an oscillating electric field, which is on the order of $1/r_0^2$. The holographic conductivity evaluated at $r_0$ should thus give us a rough idea of the response of the system to an external field with a fixed frequency of the order $1/r_0^2$. Notice that in such a picture we would expect to produce pairs if the energy is of the order of, or larger than, the mass. That is another good reason why we only explore scales much below the mass. Notice also that in the Lifshitz case of ref.~\cite{Hartnoll:2009ns}, both $\sigma(\omega,r_0)$ and $\sigma(\omega)$ had the same scaling with frequency in the limit that $\omega r^z \ll 1$.

In terms of the quantities in our formulas, for frequencies small relative to the chemical potential (the IR) we should set $r_0\gg \beta$ while for large frequencies (the UV) we should set $r_0\ll \beta$. In the low temperature limit we should send $r_H\to \infty$ relative to any other scale, and by large mass and density we mean $r_\Lambda \ll r_0$ and $\jp \gg 1/r_0^2$.

We will show in the following subsection that, with some assumptions, the D7-branes have a very simple embedding that we can compute analytically in our regimes of interest. Using our analytic results for $\theta(r)$, we will analytically compute the $\omega$-dependence of $\sigma(\omega)$ in the subsequent subsections and compare the $\omega$ scaling with the results from Lifshitz backgrounds, eq. \eqref{eq:hpstaccondresults}.

\subsection{D7-brane Embeddings}

To determine analytic forms for $\theta(r)$, we return to the DBI action of eq. \eqref{eq:generalDBIaction} and insert our ansatz for the worldvolume fields: $A_+(r)$, $A_-(r)$, and $\theta(r)$, using the background metric, NS B-field and dilaton of eqs.~\eqref{eq:schmetric}, \eqref{eq:schbfield}, and \eqref{eq:schdilaton}, respectively. The result will be precisely eq. \eqref{eq:schdccondaction}, if in eq. \eqref{eq:schdccondaction} we take $E_{\beta} = A_x'(r) = 0$. Writing $S_{D7} = - \int dr L$ (recall the text below eq.~\eqref{eq:adsd7action}), we find
\beq
-L =\frac{\N}{8r^5}\cos^3\theta(r)\sqrt{a_0 - b_0 \cos^2\theta(r) + r^2 f(r){\th'}^2} \equiv\frac{\N}{8r^5}\cos^3\theta(r)\hat{L} \,,
\eeq
where for later convenience we defined a ``reduced Lagrangian'' $\hat{L}$. Here $a_0(r)$ and $b_0(r)$ are functions of the $U(1)$ field strength, not yet evaluated on any solution for the field strength.\footnote{Alternatively, we can derive $\theta(r)$'s equation of motion by first solving for the gauge fields, plugging the results into $L$, performing a Legendre transform with respect to the gauge fields, and then finding the Euler-Lagrange equations of motion \cite{Kobayashi:2006sb}.} 
\begin{align}
\notag a_0(r) & = 1 + [1 + f(r)]r^4 A_+'(r)A_-'(r) + r^2 A_-'(r)^2\left(f(r) - \frac{r^6}{4\beta^2 r_H^4}\right)
-\frac{r^8\beta^2}{r_H^4}A_+'(r)^2 \\
b_0(r) & = r^2 f(r) A_-'(r)^2
\end{align}
The equation of motion for $\theta(r)$ is
\beq
\label{eqemb}
\left(\frac{\cos^3\theta(r) r^3 f(r)\th'}{\hat L}\right)'- \frac{\sin\theta(r)}{r^5}\left(3\cos^2\theta(r)\hat L - \frac{b_0 \cos^4\theta(r)}{\hat L}\right) = 0\,.
\eeq
Solving for the gauge fields by inverting eq.~\eqref{eq:currents}, with the charge density fixed, and plugging the solutions into $a_0(r)$ and $b_0(r)$, we find the on-shell values of $a_0(r)$ and $b_0(r)$,
\begin{align}
a_0(r) = \frac{\cos^6\theta(r) k_s(r)^2 }{\gamma^2(r) k_s(r)} &+ \frac{Q^2 r^4 f(r)}{\gamma^2(r) k_s(r)}\bigg[\beta^2 \cos^2\theta(r)  \notag \\ &- r^2{\th'}^2\left(\beta^2\frac{r^4}{r_H^4} - \beta^2\cos(2\theta(r)) + r^2\left(1 + \frac{\beta^4 \sin^4\theta(r)}{r_H^4}\right)\right)\bigg] \,,
\end{align}
\beq
b_0(r) = Q^2\frac{\beta^2 r^4 f(r)}{\gamma^2(r)}\frac{\rho(r)}{k_s(r)} \,,
\eeq
and the action evaluated on the solution is proportional to $\hat L = \cos^3\theta(r)\sqrt{k_s(r)\rho(r)}/\gamma(r)$.

As we reviewed in section \ref{sec:review}, when $A_t(r)$, or $A_+(r)$, is zero, the D7-brane can end at some value of the radial coordinate $r_\Lambda$, but when $A_t(r)$ or $A_+(r)$ is nontrivial, the D7-brane must extend all the way to the horizon, and has a spike when $r_\Lambda\ll r_H$. Along the spike and far from the horizon, $r_\Lambda < r \ll r_H$, $\theta(r)$ is approximately constant. Such a region in $r$ corresponds to scales below the mass gap of the charge carriers, where scale invariance is approximately restored.

Let us consider the limits $ r \ll r_H$ and $Q r^3\gg 1$, corresponding to low temperature and large density, as explained in the last subsection. If we take the IR limit $\beta \ll r$, we find
\beq
\gamma^2(r) = \cos^6\theta(r)\left(1 + \frac{\beta^2 \sin^2\theta(r) r^2}{r_H^4}\right) + Q^2 r^4 (r^2+\beta^2 \sin^2\theta(r)) \approx \cos^6\theta(r) + Q^2 r^6 \approx Q^2 r^6 \,.
\eeq
In the UV limit $\beta \gg r$ we find
\beq
\gamma^2(r) \approx \cos^6\theta(r) + Q^2 r^4 \beta^2 \sin^2\theta(r) \simeq Q^2\beta^2 \sin^2\theta(r) r^4 \,.
\eeq
The IR and UV limits of all the quantities that appear in the equation of motion eq. \eqref{eqemb} are similarly straightforward to determine. In addition, we can assume that $r^2{\th'}^2\ll 1$ as long as $r_\Lambda\ll r$, so we will take $1+r^2 f {\th'}^2 \approx 1$, $\sin \theta(r) \approx \sin\th_0$ and $\cos \theta(r)\approx\cos\th_0$ with $\theta_0$ a constant.

We then find that in the IR limit, to leading order, the equation of motion for $\theta(r)$ eq. \eqref{eqemb} becomes
\begin{equation}
\th'' + \frac{\beta^2\sin\th_0\cos\th_0}{r^4} = 0 \,,
\end{equation}
while in the UV limit, the equation of motion becomes
\begin{equation}
\left(\frac{\th'}{r}\right)'+ \frac{\cot\th_0}{r^3} = 0 \,.
\end{equation}
The solutions in the IR and UV limits are
\begin{align}
{\rm IR}\quad\th(r) &= \th_0 - \frac{\beta^2}{6r^2}\sin\th_0\cos\th_0, \\
{\rm UV}\quad\th(r) &= \th_0 + \frac{1}{2}\cot\th_0\log\frac{r}{r_0} \,,
\end{align}
where, following ref. \cite{Hartnoll:2009ns}, and as discussed above, we have introduced the reference scale $r_0$ obeying $r_\Lambda < r_0 \ll \beta$, which makes the argument of the logarithm in the UV solution dimensionless. Notice that in order to satisfy the condition $r^2{\th'}^2\ll 1$ in the UV solution, we need $|\cot\th_0|\ll 1$, meaning a narrow spike and probably a large mass gap $r_\Lambda\ll r_0$. The expansion is also limited to a region around the reference scale $r_0$, such that $|\log(r/r_0)| \ll 1$. Otherwise these results are consistent with all the approximations we have made.

\subsection{AC conductivity in the IR}

In the IR limit the equation of motion for the gauge field fluctuation, eq. \eqref{eq:eomflucs}, becomes simply
\beq
a_x'' + \frac{2}{r}a_x' + 4\beta^2\omega^2 a_x = 0\,,
\eeq
with solutions
\beq
a_x(r,\omega)=a_x^0 \, \frac{e^{\pm 2i \beta \omega r}}{r} \,,
\eeq
with $a_x^0$ a constant. The in-going solution corresponds to the positive sign in the exponential.

Next we follow the procedure described in ref.~\cite{Hartnoll:2009ns} for probe branes in Lifshitz geometries. As mentioned above, we will not directly apply the formula eq. \eqref{cond} in our case because the $r\to 0$ limit takes us out of the regime of our approximations. Instead we compute the local conductivity $\sigma(\omega,r_0)$ at a reference scale $r_0$ such that $\beta\omega r_0 \ll 1$. Plugging the solution into eq. \eqref{eq:localcond} and expanding the result with $\beta\omega r_0 \ll 1$, we find that at leading order the local conductivity is
\beq
\label{eq:IRcond}
\sigma(\omega,r_0)\propto\frac{\N\Qp r_0}{2\beta}\,\omega^{-1}\,.
\eeq
The $\omega^{-1}$ scaling is consistent with the result of ref. \cite{Hartnoll:2009ns}, our eq. \eqref{eq:hpstaccondresults}, for the relativistic case with dynamical exponent $z=1$.

\subsection{AC conductivity in the UV}

In the UV the equation of motion for the gauge field fluctuation eq. \eqref{eq:eomflucs} also takes a simple form. Using $\sin\th_0\simeq 1$, we find
\beq
\label{eq:uveom}
a_x'' + \frac{1}{r}a_x' + 4r^2\omega^2 a_x = 0 \,.
\eeq
The general solutions are Bessel functions. The solution describing an in-going traveling wave at the horizon is a Hankel function,
\beq
\label{eq:uvsol}
a_x(r,\omega) = a_x^0 \; H_0^{(1)}\left(\omega r^2\right) \,,
\eeq
where again $a_x^0$ is a constant. Notice that eqs. \eqref{eq:uveom} and \eqref{eq:uvsol} coincide with the equations in ref.~\cite{Hartnoll:2009ns} for a fluctuation of a probe brane's worldvolume gauge field in Lifshitz spacetime with dynamical exponent $z=2$. As before, we choose a cutoff $r_0$ such that $\omega r_0^2\ll 1$. Plugging the solution into eq. \eqref{cond} and expanding the result in powers of $\omega r_0^2$, we obtain
\beq
\label{eq:UVcond}
\sigma(\omega,r_0)\propto\frac{\N\Qp}{16}\left(\omega\log\left(\omega r_0^2\right)\right)^{-1} \,,
\eeq
which is indeed the same scaling with $\omega$ as obtained from a probe brane in a Lifshitz spacetime with $z=2$, see eq. \eqref{eq:hpstaccondresults}.

In summary: taking the temperature scale to be very low and the mass and density scales to be very high, we find that in the IR, meaning $\omega \ll \mu$, the AC conductivity exhibits relativistic scaling with frequency, while in the UV, meaning $\omega \gg \mu$, we find that the AC conductivity exhibits non-relativistic scaling with dynamical exponent $z=2$. These results clearly confirm our intuition on the bulk side, where the space is similar to AdS deep in the interior but not asymptotically, and in the field theory, where we have introduced an irrelevant deformation to $\N=4$ SYM that breaks the relativistic conformal group down to the Schr\"odinger group.

\section{Discussion and Conclusion}
\label{sec:conclusion}

Using gauge-gravity duality, we computed both DC and AC conductivities associated with a finite density of charge carriers in a strongly-coupled theory with non-relativistic symmetry. The theory was $\N=4$ SYM theory deformed by a dimension-five operator that breaks the relativistic conformal group down to the Schr\"odinger group, with dynamical scaling exponent $z=2$, and the charge carriers were comprised of a finite baryon density of massive $\N=2$ supersymmeric hypermultiplets. We found that, generally speaking, both the DC and AC conductivities exhibited relativistic scaling (with temperature or frequency, respectively) in the IR and non-relativistic scaling, with $z=2$, in the UV. These results are in accord with our expectations, given the origin of the non-relativistic symmetry via an irrelevant deformation of the theory. For the future, we can think of many questions that deserve further research.

First, a more detailed analysis of probe D-branes in Schr\"odinger spacetime would be extremely useful. Given background $Sch_5$ geometries that preserve some supersymmetry \cite{Bobev:2009mw,Donos:2009xc}, can we find supersymmetric embeddings of probe D-branes? What happens to the many phase transitions that occur in the relativistic setting when the probe D-branes are instead in $Sch_5$?

Introducing a magnetic field on the worldvolume of a probe D-brane is straightforward, and allows us to compute not only the Hall conductivity but indeed the entire conductivity tensor, in the DC limit \cite{O'Bannon:2007in,Ammon:2009jt}. What scaling does the Hall conductivity have in the Schr\"odinger case?

Moreover, with probe D-branes that do not fill all of $AdS_5$, we have more options about how to perform the NMT. Consider for example a D5-brane extended along $AdS_4\times S^2$ inside $AdS_5 \times S^5$. Here we must make a choice: the D5-brane may be extended along the T-duality direction $y$ or not. If the D5-brane is along $y$, we expect the results to be similar those of section \ref{sec:nmtwithd7} for the D7-brane. What happens when the D5-brane is transverse to $y$, however?

Perhaps the most exciting direction for future research would be model-building of the kind advocated in ref.~\cite{Hartnoll:2009ns}. Our results indicate that a straightforward way to engineer scaling exponents would be to start in the relativistic case, \textit{i.e.} probe D-branes in $AdS_5 \times S^5$, introduce a scalar to produce the desired IR exponents, and then perform the NMT. We generically expect that the result will be DC and AC conductivities that in the UV have non-relativistic scalings, with $z=2$, and in the IR have whatever scalings we initially gave them.

\section*{Acknowledgements}
We would like to thank J. Erdmenger, A. Karch, and D. Tong for reading and commenting on the manuscript, and M. Rangamani for useful discussions. A.O'B. also thanks K. Landsteiner, D. Mateos, and K. Peeters for useful discussions. C.H. additionally thanks C. Herzog, K. Jensen, and P. Meessen for useful discussions. The work of M.A. and A.O'B. is supported in part by the Cluster of Excellence ``Origin and Structure of the Universe.'' M.A. would also like to thank the Studienstiftung des deutschen Volkes for financial support. The work of C.H is supported in part by the U.S. Department of Energy under Grant No. DE-FG02-96ER40956. J.W. thanks the Galileo Galilei Institute for Theoretical Physics for hospitality and the INFN for partial support while this work was being completed. The work of J.W. is supported in part by the ``Innovations- und Kooperationsprojekt C-13'' of the ``Schweizerische Universit\"atskonferenz SUK/CRUS'' and the Swiss National Science Foundation.

\bibliographystyle{JHEP}
\bibliography{d7nonrel_v2}

\end{document}